# On the nature of nuclear dissipation, as a hallmark for collective dynamics at finite excitation [1]

by

## Helmut Hofmann [2]

Physik Department, TUM, D-85747 Garching

## Fedor A. Ivanyuk

Institute for Nuclear Research, 252028, Kiev-28

Physik Department, TUM, D-85747 Garching

and

## Shuhei Yamaji

Cyclotron Lab, Riken, Wako, Saitama, 351-01, Japan

---

[1] Supported in part by the Deutsche Forschungsgemeinschaft
[2] e-mail: hhofmann@physik.tu-muenchen.de

We study slow collective motion of isoscalar type at finite excitation. The collective variable is parameterized as a shape degree of freedom and the mean field is approximated by a deformed shell model potential. We concentrate on situations of slow motion, as guaranteed, for instance, by the presence of a strong friction force, which allows us to apply linear response theory. The prediction for nuclear dissipation of some models of internal motion are contrasted. They encompass such opposing cases as that of pure independent particle motion and the one of "collisional dominance". For the former the wall formula appears as the macroscopic limit, which is here simulated through Strutinsky smoothing procedures. It is argued that this limit hardly applies to the actual nuclear situation. The reason is found in large collisional damping present for nucleonic dynamics at finite temperature $T$. The level structure of the mean field as well as the $T$-dependence of collisional damping determine the $T$-dependence of friction. Two contributions are isolated, one coming from real transitions, the other being associated to what for infinite matter is called the "heat pole". The importance of the latter depends strongly on the level spectrum of internal motion, and thus is very different for "adiabatic" and "diabatic" situations, both belonging to different degrees of "ergodicity".

# 1  Introduction

Collective motion at finite excitation has attracted much interest in recent years. Whereas isovector modes are accessible in rather direct fashion, the only information one may obtain experimentally for isoscalar modes come from studies of the fission process. Unfortunately, for this case inferences about the collective motion itself can only be drawn in quite an indirect way. One takes experimental results on fission accompanied by emission of light particles and $\gamma$-rays and compares such data with theoretical model calculations. The latter have essentially two ingredients. First of all one needs a description of the collective motion itself by means of the Fokker-Planck or Langevin equations. This allows one to follow the dynamics along the fission path. This information has to be supplemented by a description of the emission process of the "particles". It need not be stressed that both processes are highly intertwined.

Despite such high complexity one has been able to deduce interesting results about the collective motion, as parameterized by appropriate transport coefficients. The main focus has justly been on dissipation, using whatever parameter to describe it, either by the friction coefficient itself, or by taking some combination with the inertia or the (local) stiffness of the potential. From all the many recent studies which have been published we shall only refer to [1- 4]. In [2] a compilation of data on the magnitude of dissipation has been presented. Computations of the type just mentioned have been compared with results from various microscopic models. In [3] and [4] information on the $T$-dependence of dissipation has been extracted from comparison with experimental findings.

From the seventies various theories had been developed to gain a microscopic understanding of nuclear dissipation. At first people concentrated on heavy ion collisions, but in recent years more stress has been put on nuclear fission. This is not only related to the fact that one is getting more and more data from the experimental side. As compared to the entrance phase of a heavy ion collision fission is a much slower process. For this reason the use of differential equations rather than the much more complicated integral equations is much better motivated or justified.

As can be seen from the papers mentioned previously, the theoretical models for friction give very diverging results. They sometimes differ by an order of magnitude, a feature which not only reflects the complexity of the problem, but which also hints at the strong necessity of finding its solution. In this paper we hope to be able contributing to such a goal. We want to present a study which allows one to extract relevant features of various models within one common framework. Besides on the absolute value of nuclear dissipation, much emphasis shall be put on its $T$-dependence. It is for the following theories and papers for which we will be able to establish some relationships.

  1) Let us first mention the wall formula [5]. It assumes that, independent of temperature, nucleonic motion can be described as that of independent particles which do not collide with themselves but only with the "wall" defining their boundary. An irreversible feature like friction comes about by performing a macroscopic limit within such a picture. For a recent report on all features of this model see [6], where more references may be found.

2) A similar picture is used in the approach based on linear response theory started in [7, 8]. There the "wall" is replaced by a deformed shell model potential, the nucleons move in quantum states and are allowed to "scatter" from one another. Details of this theory can be found in [9- 13], (see also [14]), together with numerical computations of transport coefficients.

3) Finally, we like to mention theories for which friction shows a "hydrodynamical" behavior, in the sense of being proportional to a relaxation time $\tau_{\text{intr}}$ of nucleonic motion, and thus to $T^{-2}$.

   3.1) There is the theory of "dissipative, diabatic dynamics" (DDD) proposed in [15] (for a review see [16]), which bases on the assumption that nuclear collective motion happens predominantly "diabatically". Otherwise, a similar picture is used as the one mentioned earlier, in the sense that nucleons move in shell model potentials, but with relaxation processes being taken into account explicitly. The non-Markovian features, which this model may also account for to describe the entrance phase of heavy ion collisions, will not be considered here. Rather, we want to concentrate on its limit for slow motion.

   3.2) For such a situation a similar result was derived in [17]. There the von Neumann equation had been applied to the deformed shell model, complemented by a collision term in relaxation time approximation.

   3.3) For the two previous models the association to hydrodynamics is only given somewhat loosely through the proportionality factor $\tau_{\text{intr}}$ in the friction coefficient, or of components of it. Hydrodynamical viscosity in proper sense of "collisional dominance" is found whenever the nucleonic dynamics is described by transport equations like the Landau equation with collision term. We just like to mention one recent work [18], which combines the use of such an equation with a special treatment of the surface by way of collective variables.

   3.4) In [19] a model has been presented in which collective dynamics itself is governed by two body collisions, rather than by the picture of a time dependent mean field. This implies "collisional dominance" by its very construction. It is therefore not very astonishing that in the end one deduces a friction coefficient which decreases with temperature as $T^{-2}$. Such a behavior is found although at intermediate steps no reference is being made to common concepts of fluid mechanics such as relaxation times etc.

As we have seen, these models spread over the whole range of assumptions one may make for nuclear dynamics, from the *pure* independent particle model to the ones which are entirely governed by collisions. It seems clear, therefore, that an understanding of nuclear dissipation will greatly help to understand better the general nature of collective motion at finite temperatures. Perhaps the present work may contribute to this goal, as we will be able to encompass within one single model all the limits of dissipation mentioned.

The paper is organized as follows. In chapter 2 we shall review linear response theory. In the first part some details of the general theory will briefly be reported from previous publications. In the second part we will present a thorough discussion of some specific thermal properties of nucleonic motion, those properties which in the end will help us to build the bridge between the opposite models of friction mentioned. This comparison will finally be completed in chapter 5 on the basis of numerical calculations. In chapter 3 we shall discuss the model of DDD following [16], adding some remarks about [17] and [18]. Chapter 4 will be devoted to a derivation of the wall formula by means of Strutinsky smoothing procedures.

Throughout the approach we will make the assumption of collective motion being sufficiently slow such that large scale motion can be linearized locally. As we will be concentrating on average motion, discarding any fluctuations, the only condition which needs to be fulfilled is the one of having the collective time scale much larger than the microscopic one. Such a situation should be given for fission, at least if the barrier is not too small, as compared to temperature. In such a case it will take a long time before on average the system moves across the barrier. Anticipating strong frictional forces, the situation will favor this linearization scheme even behind the barrier. In any case the typical time scale for collective motion can be expected to be larger than or of the order of some few $10^{-21}$s, and thus bigger by about one order of magnitude than the time typical for nucleonic motion .

## 2 Linear response theory for collective motion

In this section we briefly outline the application of response theory, following largely the presentation in [20] (see also [21] for the case of $T = 0$). We take for granted to be given a Hamiltonian $\hat{H}(\hat{x}_i, \hat{p}_i, Q)$ for the nucleons' dynamics in a deformed mean field, with the deformation being parameterized by the shape variable $Q$, whose average $\langle \hat{H}(\hat{x}_i, \hat{p}_i, Q) \rangle$ represents the total energy of the system $E_{\text{tot}}$ (eventually *including* both the Strutinsky re-normalization as well as "heat"). The equation of motion (EOM) for $Q(t)$ can then be constructed from energy conservation. From Ehrenfest's equation it follows:

$$0 = \frac{d}{dt} E_{\text{tot}} = \dot{Q} \left\langle \frac{\partial \hat{H}(\hat{x}_i, \hat{p}_i, Q)}{\partial Q} \right\rangle_t \equiv \dot{Q} \left\langle \hat{F}(\hat{x}_i, \hat{p}_i, Q) \right\rangle_t \qquad (2.1)$$

All one needs to do to get the equation of motion for $Q(t)$ is to express the average $\left\langle \hat{F}(\hat{x}_i, \hat{p}_i, Q) \right\rangle_t$ as a functional of $Q(t)$. The operator which matters is seen to be given by the derivative of the mean field with respect to $Q$. Assuming "collisions" to act fairly independent of deformation this operator $\hat{F}(\hat{x}_i, \hat{p}_i, Q)$ is of one body nature.

## 2.1 Intrinsic versus collective response

Provided collective motion is sufficiently slow the EOM can be obtained by linearizing locally in $Q$. To this end one may expand the $\hat{H}(Q)$ around some $Q_0$ to give:

$$\hat{H}(Q) = \hat{H}(Q_0) + (Q - Q_0)\hat{F} + \frac{1}{2}(Q - Q_0)^2 \left\langle \frac{\partial^2 \hat{H}}{\partial Q^2}(Q_0) \right\rangle_{Q_0, T_0}^{\mathrm{qs}} \tag{2.2}$$

Here and in the sequel the $\hat{F}(\hat{x}_i, \hat{p}_i, Q)$ shall be denoted by $\hat{F}$ whenever it is to be taken at $Q_0$. A lengthy derivation then leads from (1) to the following form of the local EOM

$$k^{-1} q(t) + \int_{-\infty}^{\infty} \tilde{\chi}(t-s) q(s) ds = 0 \tag{2.3}$$

Here $q = Q - Q_m$ measures the deviation of the actual $Q$ from the center of the oscillator approximating the true potential in the neighborhood of $Q_0$. The $\tilde{\chi}$ is the causal response function associated to the dynamics of the nuclear "property" $\langle \hat{F} \rangle$. It is given by

$$\tilde{\chi}(t-s) = \Theta(t-s)\frac{i}{\hbar}\mathrm{tr}\left(\hat{\rho}_{\mathrm{qs}}(Q_0, T_0)[\hat{F}^I(t), \hat{F}^I(s)]\right) \equiv 2i\Theta(t-s)\tilde{\chi}''(t-s) \tag{2.4}$$

with the time evolution in $\hat{F}^I(t)$ as well as in the density operator $\hat{\rho}_{\mathrm{qs}}$ being determined by $H(Q_0)$. The $\hat{\rho}_{\mathrm{qs}}$ is meant to represent thermal equilibrium at $Q_0$ with excitation being parameterized by temperature or by entropy. The quantity $k$ summarizes contributions of static forces which appear in second order. Anticipating the changes in entropy to be quadratic in $\dot{q}(t)$—and hence beyond the order considered when deriving (3)—one gets for the coupling constant $k$ (see [20]):

$$-k^{-1} = \left\langle \frac{\partial^2 \hat{H}}{\partial Q^2}(Q_0) \right\rangle_{Q_0, T_0}^{\mathrm{qs}} + (\chi(0) - \chi^{\mathrm{ad}}) \tag{2.5}$$

with $\chi(0)$ being the static response and $\chi^{\mathrm{ad}}$ the adiabatic susceptibility. Below we will show a form more suitable for numerical applications.

To make these formulas plausible a few explanations are in order. It is easily seen that (3) has the same structure as the equation of motion which conventionally describes collective vibrations. Here it is applied also for cases when $Q_m$ does not coincide with a minimum of the potential surface. More important, these formulas are correct at finite excitations and under the presence of collisions. The form (5) is a generalization of the one given in [21] (which can be seen to be related to the one of Bohr-Mottelson). The last term guarantees that to harmonic order in $q$ the entropy $S$ of the system is conserved, for which reason the adiabatic susceptibility occurs. It is defined by the relation $\delta \langle \hat{F} \rangle \big|_{Q_0, S} = -\chi^{\text{ad}} \delta q$, the difference to the static response being that for the latter a similar relation holds true but without specifying to constant entropy.

Fourier transforming (3) leads to the secular equation

$$\chi(\omega) + k^{-1} = 0 \tag{2.6}$$

for the possible excitations of the system. It is convenient to have the latter appear as poles of another function, the one which measures the response to an "external" field. It is only in this response function that collective modes appear which are associated to the field $\hat{F}$. Since they even dominate the strength distribution we like to call this response function the "collective one", $\chi_{\text{coll}}(\omega)$. It can be constructed by adding to $\hat{H}(\hat{x}_i, \hat{p}_i, Q)$ a coupling $f_{\text{ext}}(t)\hat{F}$, i.e. by introducing a Hamiltonian $\hat{H}' = \hat{H}(\hat{x}_i, \hat{p}_i, Q) + f_{\text{ext}}(t)\hat{F}$. The $\chi_{\text{coll}}(\omega)$ can then be defined in the usual way, namely by $\delta \langle \hat{F} \rangle_\omega = -\chi_{\text{coll}}(\omega) f_{\text{ext}}(\omega)$. For its derivation one may follow closely the one known for zero temperature. It becomes completely identical to the latter provided the internal degrees of the system behave "ergodic", in the sense that the static response $\chi(0)$ is identical to the adiabatic susceptibility $\chi^{\text{ad}}$ [22]:

$$\chi(0) = \chi^{\text{ad}} \tag{2.7}$$

In this case one gets:

$$\chi_{\text{coll}}(\omega) = \frac{\chi(\omega)}{1 + k\chi(\omega)} \tag{2.8}$$

Henceforth we are going to assume (7) to be fulfilled, and we will come back to its physical implications below. For lack of space we do not want to touch upon the relevance the condition (7) would have on the time evolution of our system. The notion "ergodic" shall thus only be used to designate the property (7).

## 2.2 Collisional damping of nucleonic motion

So far we have not specified the Hamiltonian $\hat{H}(Q_0)$. There is no doubt that it should contain an interaction $\hat{V}_{\text{res}}^{(2)}(\hat{x}_i, \hat{p}_i)$ *residual* to the mean field. It is this interaction which in the end is responsible for the damping mechanism. We want to assume this interaction not to depend on $Q$, implying that the $\hat{F}$ is a pure one body operator. Inspecting (8) it becomes apparent that this interaction enters the game through the *nucleonic* response function $\chi(\omega)$. In an ideal calculation one would like to evaluate the latter from the basic definition (4) using the spectrum and the eigenstates of the Hamiltonian $\hat{H}(Q_0)$,

$$\hat{H}(Q_0) \mid n(Q_0) \rangle = E_n(Q_0) \mid n(Q_0) \rangle \tag{2.9}$$

It is easy to prove that the dissipative part of the response function takes on the form (see e.g. [7, 23])

$$\chi''(\omega) = \pi \sum_{nm} \rho(E_m) \left| F_{mn} \right|^2 \left[ \delta\left(\omega - (E_n - E_m)\right) - \delta\left(\omega + (E_n - E_m)\right) \right] \quad (2.10)$$

with the matrix elements $F_{mn} = \langle m|\hat{F}|n\rangle$.

Generally, an expression like (10) can be calculated exactly only within the pure single particle model. For a finite $\hat{V}_{\text{res}}^{(2)}(\hat{x}_i, \hat{p}_i)$ approximations are necessary. We want to treat it in some analogy to the way one expects "collisions" to modify single particle motion. This is possible by applying the technique of Green functions (see [23]), following the early suggestion in [9] (see also [24]) which has been applied in various computations since (see e.g. [11, 12, 14]). For detailed descriptions we like to refer to [25, 26, 12] and [23]. In this paper we want to explain the method in more heuristic way, starting by evaluating the response function in the pure single particle picture. Writing the field in second quantization

$$\hat{F}(t) = \sum_{jk} F_{jk} \hat{c}_j^\dagger(t) \hat{c}_k(t) \quad (2.11)$$

a straightforward calculation leads to

$$\tilde{\chi}''(t) = \sum_{jkj'k'} F_{jk} F_{k'j'} \frac{1}{2\hbar} \langle \left[ \hat{c}_j^\dagger(t)\hat{c}_k(t), \hat{c}_{k'}^\dagger \hat{c}_{j'} \right] \rangle \quad (2.12)$$

for the dissipative part of the response function. Here, the $|k\rangle$ are the eigenstates of the single particle Hamiltonians $\hat{h}(\hat{x}_k, \hat{p}_k, Q_0)$ (with corresponding energies $e_k$), which constitute the $\hat{H}(\hat{x}_i, \hat{p}_i, Q_0)$ when summed over all particles. The expectation value appearing on the right hand side of (12) is easily seen to be diagonal in $k = k', j = j'$, so that the Fourier transform of (12) can be written as

$$\chi''(\omega) = \sum_{jk} |F_{jk}|^2 \chi''_{jk}(\omega) \quad (2.13)$$

with

$$\chi''_{jk}(\omega) = \frac{1}{2} \int_{-\infty}^{\infty} \frac{d\Omega}{2\pi\hbar} \left( n(\Omega - \frac{\omega}{2}) - n(\Omega + \frac{\omega}{2}) \right) \varrho_k(\Omega - \frac{\omega}{2}) \varrho_j(\Omega + \frac{\omega}{2}) \quad (2.14)$$

and the $\varrho_k(\omega)$ being given by $\varrho_k(\omega) = 2\pi \delta(\hbar\omega - e_k)$. Of course, for this simplified case we easily could have integrated over $\Omega$ to get a much simpler expression. But (13, 14) retain their validity under more general conditions.

Notice that the $\varrho_k(\omega)$ represents the strength with which the single particle state $|k\rangle$ contributes to $\chi''_{jk}(\omega)$. But under the presence of residual couplings this strength distributes over more complicated states. This feature may be parameterized by means of the real and imaginary parts of a self-energy $\Sigma(\omega \pm i\epsilon) = \Sigma'(\omega) \mp \frac{i}{2}\Gamma(\omega)$ to give for $\varrho_k(\omega)$

$$\varrho_k(\omega) = \frac{\Gamma(\omega)}{(\hbar\omega - e_k - \Sigma'(\omega))^2 + \left(\frac{\Gamma(\omega)}{2}\right)^2} \qquad (2.15)$$

Although this form correctly represents the strength of the state $\mid k \rangle$, it should not be concealed that (13), together with (14) and (15), is only an approximation to $\chi''(\omega)$. The physical argument we have inherently used is the one of *statistical independence*: Since we are calculating the *nucleonic* response function we may assume the individual excitations contributing to the full expression (see (12)) *not to be correlated*. Notice, please, that by construction *all collective modes* are supposed to be treated *explicitly* by collective variables like the one $Q$ we chose to concentrate on in this paper.

To specify (15) fully it is necessary to have a model for the self-energies. In the references [25], [27] the following form has been suggested for $\Gamma(\omega, T)$ of the imaginary part:

$$\Gamma(\omega, T) = \frac{1}{\Gamma_0} \frac{(\hbar\omega - \mu)^2 + \pi^2 T^2}{1 + \frac{1}{c^2}[(\hbar\omega - \mu)^2 + \pi^2 T^2]} \qquad (2.16)$$

with $\mu$ being the chemical potential. The real part $\Sigma'(\omega)$ is obtained by a Kramers-Kronig relation. The $1/\Gamma_0$ represents the strength of the "collisions", viz of the coupling to more complicated states. The cut-off parameter $c$ allows one to account for the fact that the imaginary part of the self-energy does not increase indefinitely when the excitations get away from the Fermi surface. Both parameters are not known precisely, but from experience with the optical potential and the effective masses [28, 29] the following range of values can be given

$$\begin{aligned} 0.03 \quad MeV^{-1} &\leq \quad \Gamma_0^{-1} \quad \leq \quad 0.06 \quad MeV^{-1} \\ 15 \quad MeV &\leq \quad c \quad \leq \quad 30 \quad MeV \end{aligned} \qquad (2.17)$$

Neglecting the $\omega$ dependence of $\Gamma$ and putting $c \to \infty$ the values given in (17) leads to a average relaxation time for single particle motion $\tau_{\text{int}} = \hbar/\Gamma$ which is in accord with the estimate given in [30].

The alert reader will have recognized that for small excitations the form (16) reduces to the expression well known from Fermi liquid theory. The presence of the parameter $c$ reduces the strong dependence on both $(\hbar\omega - \mu)$ and $T$ at larger excitations. However, this reduction certainly is not big enough to diminish the size of the collisional width $\Gamma(\omega)$ sufficiently strongly at very large $(\hbar\omega - \mu)$ or very large $\pi T$. This may cause problems in actual computations if one were to apply the model at too large temperatures, or if one would look at frequencies too far away from the Fermi surface. Then the $\omega$-dependence given by (16) can no longer be trusted. The conclusion to draw from this observation is to either leave out such contributions entirely or to treat them by decent regularization schemes. With respect to the frequency dependence, fortunately, the only place where such a problem occurs is in the evaluation of sum rules. Then this regularization scheme should be worked out in such a way as to guarantee the sum rule to be fulfilled.

Finally we should like to point to another simplification used in (16). The form chosen there for the imaginary part does not lift degeneracies present in the single particle model. To achieve such goal by way of our self-energies one would have to make the $\Gamma(\omega, T)$ depend on the quantum numbers of the single particle states—in addition to the dependence on the energy (or frequency) taken into account. To study this feature in full glory one has to calculate from (16) the real parts of the self-energies through a Kramers-Kronig relation (see e.g. [23,25]). But for the present paper we would not like to consider such a refinement, although we will have to face consequences later on.

## 2.3 The effective coupling constant

The derivation of the EOM in sect.2.1 was based on the linearization about a thermal equilibrium. As the coupling constant $k$ of (5) is entirely determined by quasi-static properties, it is no surprise that formulas can be derived which involve either the internal energy $E(Q, S_0)$ at given entropy $S_0$ or the free energy at given temperature $T_0$ [20] (even without assuming (7)). The first one reads:

$$-k^{-1} = \left.\frac{\partial^2 E(Q, S_0)}{\partial Q^2}\right|_{Q_0} + \chi(0) \equiv C(0) + \chi(0) \qquad (2.18)$$

In this quasi-static picture the $E(Q, S_0)$ stands for the lowest possible energy, and, hence, relates to what often has been associated with "adiabatic dynamics", in particular at zero excitation. The opposite case is the one of "diabatic motion". Taking this model to the extreme, the occupation numbers of the nuclear states are frozen. Interestingly enough, for vibrations of this type the previous formulas can be taken over. The only change necessary is to replace $k$ by a *diabatic coupling constant* $k_{\rm di}$ whose value is given by an expression like (18) with the quasi-static energy being replaced by the diabatic one $E_{\rm di}(Q)$, see [20]. One gets

$$-k_{\rm di}^{-1} = \left.\frac{\partial^2 E_{\rm di}(Q)}{\partial Q^2}\right|_{Q_0} + \chi(0) \qquad \Longrightarrow \qquad -k_{\rm di}^{-1} + k^{-1} = \left.\frac{\partial^2 E_{\rm di}(Q)}{\partial Q^2}\right|_{Q_0} - \left.\frac{\partial^2 E(Q, S_0)}{\partial Q^2}\right|_{Q_0} \qquad (2.19)$$

with the $E_{\rm di}(Q)$ representing the diabatic energy surface. As discussed in [20] the $k_{\rm di}$ stays constant with excitation, different to the case of "adiabatic" motion whose coupling constant strongly decreases with $T$ (see Fig.5 of [20]).

## 2.4 Strength distributions

The dissipative part of $\chi_{\text{coll}}(\omega)$ as given by (8) represents the distribution of strength over the various possible modes. As we shall see, this distribution behaves completely different for "adiabatic" and "diabatic" motion.

In [14] isoscalar quadrupole vibrations of $^{208}$Pb around the stable spherical configuration have been studied. At small temperatures ($T \sim 0.5$ MeV) the strength distribution looks similar to the one known from $T = 0$. But above some critical value of $T \sim 2-3$ MeV the high lying modes disappear and almost all strength concentrates in a broad peak at a very small frequency. According to [20] this is largely due to the fact that the vibrations looked at are the ones about thermal equilibrium in true sense, meaning that many-particle, many-hole configurations come into play, for instance through the coupling constant (18).

Still according to [20] a completely different behavior is to be expected if the coupling constant is evaluated according to (19) in the "extreme diabatic" picture with all the occupation numbers frozen. In Fig.2.1 we present a numerical computation, again for the same situation as in [14], but with $k$ replaced by $k_{\text{di}}$. As expected, no shift of strength is seen. It is only that the giant peaks get broader at larger $T$, an effect which easily is traced back to the $T$-dependence of the (bare) single particle width as given by (16).

## 2.5 Transport coefficients and their dependence on $T$

In general, the strength distribution shows individual peaks. They may be interpreted to represent individual modes ("resonances") of the system. For each one we may then define transport coefficients for average motion, namely $M, \gamma, C$ by identifying for the corresponding range of frequencies an oscillator response function through

$$\left(\chi_{\text{coll}}(\omega)\right)^{-1} \delta\langle\hat{F}\rangle_\omega \simeq \left(\chi_{\text{osc}}(\omega)\right)^{-1} \delta\langle\hat{F}\rangle_\omega \equiv \left(-M\omega^2 - \gamma i\omega + C\right) \delta\langle\hat{F}\rangle_\omega = -f_{\text{ext}}(\omega) \tag{2.20}$$

The behavior of the strength distribution with increasing $T$, as seen in the "adiabatic case", must have implications on the transport coefficients. For instance, it is well known that any mode whose strength exhausts the energy weighted sum will have an inertia close to the one of irrotational flow $m_0$. Indeed, in [14] it was found that the inertia of the low frequency mode, the only one which survives at large $T$, does turn into $m_0$ for $T \geq 2-3$ MeV. This is a clear indication that the transition discussed is related to one of a macroscopic limit. For friction $\gamma$ the situation is less evident. In [12] it was found that $\gamma$ increases with $T$ (for $T \leq 4$ MeV)—a behavior which is entirely different both from the one of hydrodynamics as well as from the one of the wall formula. We hope to be able help clarify this question in this paper.

For the quadrupole vibrations of $^{208}$Pb all transport coefficients have been computed. Their temperature dependence shall be summarized looking at parameters discussed in the literature:

| $T =$ | 0.5 | 1.0 | 2.0 | 3.0 | 4.0 | ( MeV) |
|---|---|---|---|---|---|---|
| $\eta = \frac{1}{2}\gamma/\sqrt{M\mid C\mid}$ | 0.07 | 0.46 | 1.62 | 2.8 | 4.6 | dim.less |
| $\beta = \gamma/M$ | 0.56 | 3.0 | 5.8 | 9.6 | 11.5 | ($10^{21}$ s$^{-1}$) |
| $\tau_{\text{coll}} = \gamma/\mid C\mid$ | 0.03 | 0.29 | 1.80 | 3.3 | 7.50 | ($10^{-21}$ s) |
| $\hbar\varpi = \hbar\sqrt{\mid C\mid/M}$ | 2.8 | 2.1 | 1.19 | 1.12 | 0.81 | ( MeV) |

Computations along the fission path are under way. Preliminary results show a very similar behavior of the transport coefficients with temperature. No surprise is seen for the dependence on $Q$: For $T = 1$ MeV the coefficient $\beta$, for instance, only changes by a factor of 2 when calculated along the fission path. It actually *decreases* with increasing elongation, like it is the case for the friction coefficient itself. Details will be published in [31]. It must be said that in these computations the influence of diagonal matrix elements to the nucleonic response has been neglected. It will be one of the top issues of the discussion to come below to clarify the physical significance of such a constraint.

In the table there is included the quantity $\tau_{\text{coll}}$ which becomes the relevant time scale for strongly overdamped motion, for which one has

$$\beta^{-1} = \frac{M}{\gamma} \equiv \tau_{\text{kin}} \quad \ll \quad \tau_{\text{coll}} \equiv \frac{\gamma}{\mid C\mid} \qquad (2.21)$$

From $\tau_{\text{coll}}/\tau_{\text{kin}} = 4\eta^2$ one sees (21) to be fulfilled for $4\eta^2 \gg 1$, which according to the table should be given for temperatures above $T \approx 2$ MeV. In this limit the inertia $M$ drops out of the equations of motion. Consequently, both the parameter $\eta$ as well as the $\beta$ become irrelevant. Physically, the $\tau_{\text{kin}}$ represents the time in which the collective kinetic energy relaxes to the Maxwell distribution. The remaining parameter determines the time scale for the creeping motion along the potential landscape.

### 2.5.1 Friction coefficient in zero frequency limit

The definition of the transport coefficients given above involves both the computation of the collective response function as well as its analysis in terms of selected peaks of the strength distribution and their interpretation as possible modes. For very slow modes it is possible to deduce the secular equation typical for the oscillator (see the part on the right of (20)) directly from the nucleonic response. Generally speaking this is possible whenever the low frequency poles of (8) can be found by expanding $\chi(\omega)$ to second order in $\omega$. This is to write

$$\left(\frac{1}{k} + \chi(0)\right) + \omega \left(\frac{\partial \chi}{\partial \omega}\right)_{\omega=0} + \omega^2 \left(\frac{1}{2}\frac{\partial^2 \chi}{\partial \omega^2}\right)_{\omega=0} = 0 \qquad (2.22)$$

from which equation the transport coefficients $M(0), \gamma(0), C(0)$ of the "zero frequency limit" are easily recognized, just by comparing with the form (20). For the friction coefficient we thus get

$$\gamma(0) = -i \frac{\partial \chi(\omega)}{\partial \omega} \bigg|_{\omega=0} = \frac{\partial \chi''(\omega)}{\partial \omega} \bigg|_{\omega=0} \qquad (2.23)$$

and after inserting (13)-(14):

$$\gamma(0) = -\int \frac{d\hbar\Omega}{4\pi} \frac{\partial n(\Omega)}{\partial \Omega} \sum_{jk} |F_{jk}|^2 \varrho_k(\Omega) \varrho_j(\Omega). \qquad (2.24)$$

An essential difference between the two versions of introducing transport coefficients is found in the following fact: The former, more general definition naturally accounts for non-Markovian effects (see [32]) and it warrants self-consistency. We know from experience that it is mostly for the inertia that both versions differ considerably. To reasonably good approximation we may take over the zero frequency limit for friction, on which we want to concentrate later on.

## 2.6 The role of symmetries

In this section we are going to examine more closely the influence the nucleonic spectrum has on collective properties. Of particular interest will be the question of degeneracies. So far we have not specified whether or not the sums over $j, k$ appearing in the response function (13), and hence in the friction coefficient (24), should include matrix elements with $e_j = e_k$. Quite generally, one expects the latter to contribute to quasi-static properties of the system and in this way eventually to conservative forces. As for friction, finite contributions to $\gamma(0)$ can only come from transitions between micro-states $|n\rangle$ and $|m\rangle$ if their energies are different, $E_n \neq E_m$. This behavior is intuitively clear, and it is not too difficult to prove it formally starting from the microscopic form (10) of the response function (for details see [23]). Translating to (24), however, this feature does not necessarily imply to exclude contributions to the sum from terms with $e_k = e_j$. Indeed, as soon as the single particle states have some width, as given by the $\Gamma(\omega, T)$ of (16), for instance, the distributions $\rho_k(\Omega)$ encompass a whole spectrum of such micro-states.

In this section we are going to examine the relevance of such contributions. First we will look at general quasi-static properties, to study the implications on dissipation afterwards. This will involve some discussion about ergodicity. Interestingly enough, it is here where we will see differences between "diabatic" or "adiabatic" level schemes.

### 2.6.1 Quasi-static properties and the "heat pole"

Probing the system to linear order in the $Q - Q_0$ its static properties can be parameterized in terms of susceptibilities, which we are now going to study.

## a) Susceptibilities

We already have mentioned two of them, the adiabatic one and the static response, which sometimes is called isolated susceptibility [33, 22]. Besides them there is the isothermal one for which the change in $\langle \hat{F} \rangle$ is to be calculated at fixed temperature. This quantity would indeed appear in the coupling constant $k^{-1}$ in case that locally the collective motion would happen at constant $T$. Then the $k^{-1}$ were given by a formula like (18) but with the internal energy replaced by the free energy. As shown in [20], the difference between both can be written as

$$-k^{-1} + k^{-1}\big|_{T=\text{const.}} = \chi^{\text{T}} - \chi^{\text{ad}} = -\left(\frac{1}{\frac{\partial^2 f}{\partial T^2}}\left(\frac{\partial^2 f}{\partial T \partial Q}\right)^2\right)_{Q_0,T} = \frac{1}{T}\left(\frac{\langle \delta\hat{F}\delta\hat{H}\rangle\langle\delta\hat{H}\delta\hat{F}\rangle}{\langle\delta\hat{H}^2\rangle}\right) \quad (2.25)$$

(like (18) also this result is correct independently of (7)). The difference in the two coupling constants can be understood as a measure to which degree the approximate concept of a *constant* temperature will be fulfilled in realistic situations. Numerical estimates have been presented in [34] and in [20]. It was found that the difference $\chi^{\text{T}} - \chi^{\text{ad}}$ is negligible small for temperatures above $T \approx 1.5$ MeV, the range we are interested in the present paper. We may add in passing that the mixed derivative of the free energy will be exactly zero at equilibrium positions which do not change with $T$.

Finally, we aim at calculating the difference $\chi(0) - \chi^{\text{ad}}$. Formally this quantity can be expressed [33, 35] as a sum over fluctuations of the type $\langle\hat{\delta}\hat{F}\delta\hat{H}_\nu\rangle\langle\hat{\delta}\hat{H}_\nu\delta\hat{F}\rangle/\langle\hat{\delta}\hat{H}_\nu^2\rangle$ with the $H_\nu$ representing all possible constants of motion, including powers of the Hamiltonian itself. But this result is too complicated to allow for numerical evaluations in the general case. Therefore, we want to go a small detour by first looking at $\chi^{\text{T}} - \chi(0)$, for which the following form can be derived (see Append.A of [20])

$$\chi^{\text{T}} - \chi(0) = \frac{1}{T}\sum_{\substack{n,m \\ E_n=E_m}} \langle m|\delta\hat{F}|n\rangle\langle n|\delta\hat{F}|m\rangle \rho(E_m) = \frac{1}{T}\langle\delta\hat{F}^0\delta\hat{F}^0\rangle \quad (2.26)$$

In the expression in the middle there appear the eigenstates and energies of the Hamiltonian, as given by (9). The $\delta\hat{F}^0$ on the very right stands for the so called "zero frequency component" of the operator $\delta\hat{F}$ [33, 22]. In our notation this component is obtained by evaluating the Fourier transform of $\hat{F}(t) = \exp(i\hat{H}t/\hbar)\hat{F}\exp(-i\hat{H}t/\hbar)$ at $\omega = 0$. Using the spectral representation of $\hat{H}$, it is easily seen to be given by

$$\hat{F}^0 = \sum_{\substack{n,m \\ E_n=E_m}} \langle m|\hat{F}|n\rangle |m\rangle\langle n| \quad (2.27)$$

This $\hat{F}^0$ commutes with the Hamiltonian.

Under certain conditions the expression on the very right of (26) becomes identical to the one on the very right of (25), such that (7) is fulfilled. In [22] a system having such properties is called ergodic. As demonstrated in chap.3 of [22] (cf. also [35]), it is sufficient to have a

i) a non-degenerate spectrum $E_m$

ii) a narrow distribution of the occupied states.

The first condition says that all constants of motion $\hat{H}_\nu$ must commute with each other, implying that they can be expressed in terms of powers $\hat{H}^k$. The second one comes in because the sum involving the fluctuations $\delta\hat{H}_\nu$ mentioned above then reduces to an expansion in terms of powers of fluctuations in energy.

The important question arises to which extent we may assume these conditions to be given in the nuclear case, in particular for the situation we are faced with studying collective motion whose generator is the one body operator $\hat{F}$ for which these susceptibilities are to be evaluated. As for (i), certainly *there are a few* conserved quantities which do not commute with each other, like the components of the total angular momentum $\hat{\mathbf{J}}$, or eventually the projection $K$ of $\hat{\mathbf{J}}$ on an axis of symmetry. Thus the condition we speak of can at best be fulfilled within the subspaces of given total angular momentum. However, a decent projection on angular momentum is too difficult to be performed for applications we have in mind for our theory, namely large scale collective dynamics of heavy nuclear systems. Rather, one commonly describes this dynamics in the so called body fixed system without caring about angular momentum conservation. For such a situation the assumption of a non-degenerate system makes sense, at least if we refer to complex many-body states beyond the pure independent particle model.

Indeed, it is experimentally established fact that the *states of the compound nucleus are non-degenerate*, to the exception of accounting properly for the few conserved quantities mention before. The characteristic distribution of levels as function of their mutual distance is of Wigner type rather than Poisson, a feature which is associated to chaotic behavior of nuclear dynamics [36]. We take it as clear evidence that one of the two basic conditions needed for ergodicity is indeed given for nuclear dynamics, in the sense described above. Of course, this fact hinges strongly on the effects of residual interactions.

On the level of the mean field the question of degeneracies is closely related to the distinction between the "adiabatic" or "diabatic" model. In a deformed shell model the single particle levels cross at many places, at which the levels become degenerate. Lifting this degeneracy by considering some interaction, the "diabatic" spectrum turns into the "adiabatic" one, for which two levels may come close but never cross. It is this situation which we should like to favor in order to ensure ergodicity, in the sense of having (7) fulfilled. In practical applications, deformed shell models will exhibit more level crossings the higher the symmetry of the model will be. Along the path to fission the system can be expected to reach complex shapes such that the single particle levels can be considered largely non-degenerate.

So far we have only been discussing condition (i). Unfortunately, with respect to the second one (ii) the nuclear situation is somewhat less favorite. Generally speaking, the nucleus is too small for the canonical distribution to become as negligible narrow as for a truly macroscopic system. An adequate distribution for a nucleus would thus be the micro-canonical one, whose width can be adjusted to an experimental situation. For many applications the introduction of temperature still makes sense. The price to pay is found in an uncertainty $\Delta T$ of temperature $T$. Whenever the quantities of interest do not change with $T$ too strongly the error one makes is well under control. But the situation becomes entirely different for quantities which depend on the fluctuations of the energy themselves * —as it is the case for $\chi^{\rm ad} - \chi(0)$, which, as mentioned, can be expressed as an expansion in powers of exactly those fluctuations. These facts seem to indicate that this is a place where special measures are in order to cure for such deficiency—if for pragmatic reasons one wants to stick to the concept of temperature elsewhere. A practical method will be discussed below.

**b) Heat pole**

In the previous discussion it was seen that the quantity $\chi^{\rm T} - \chi(0)$ plays a crucial role when we try to understand the question of ergodicity, last but not least because it is this difference which is accessible to numerical computations. This is true in particular when we want to examine this problem in the framework of our theory in which we account for the residual interaction $\hat{V}_{\rm res}^{(2)}(\hat{x}_i, \hat{p}_i)$ by way of collisional damping (see *sect.*{2.2}). These features can be studied better with a correlation function rather than the response function. The symmetrized version of the former is defined by $\tilde{\psi}''(t)$

$$\tilde{\psi}''(t-s) = \frac{1}{2}{\rm tr}\hat{\rho}_{\rm qs}[\hat{F}(t) - \langle\hat{F}\rangle, \hat{F}(s) - \langle\hat{F}\rangle]_+ \qquad \langle\hat{F}\rangle = {\rm tr}\hat{\rho}_{\rm qs}\hat{F} \qquad (2.28)$$

Analyzing its Fourier transform in terms of an exact spectral representation with respect to the total Hamiltonian, one realizes that the $\psi''(\omega)$ has a $\delta$ function type singularity at $\omega = 0$. This is to say that the full function can be split like

$$\psi''(\omega) = \psi^0 2\pi\delta(\omega) + {}_R\psi''(\omega) \qquad (2.29)$$

with the ${}_R\psi''(\omega)$ being regular at $\omega = 0$. Here, the pre-factor $\psi^0$ of this $\delta$ function can be seen to be given by the expectation value $\langle\delta\hat{F}^0\delta\hat{F}^0\rangle$ of the squared zero-frequency part of $\delta\hat{F}$, namely

$$\psi^0 = \langle\delta\hat{F}^0\delta\hat{F}^0\rangle = T\left(\chi^{\rm T} - \chi(0)\right) \qquad (2.30)$$

---

* As noted in [37] (c.f. comment below eq.(28,8)), it is especially this property for which the canonical distribution must not be used for an isolated system .

with the last equation following from (26). In the sequel we like to call this singularity the "heat pole" with the $\psi^0$ being its residue.

Let us see how the correlation function $\psi''(\omega)$ looks like in our model for collisional damping. We may apply the same procedure as in the case of the response function. One gets a form like (2.13), but with the $\chi''_{jk}(\omega)$ of (2.14) replaced by

$$\psi''_{jk}(\omega) = \pi \int_{-\infty}^{\infty} \frac{d\Omega}{2\pi} \frac{d\Omega'}{2\pi} \, n(\hbar\Omega)\,(1 - n(\hbar\Omega'))\, \varrho_k(\Omega')\varrho_j(\Omega) \times \\ \left(\delta(\omega - \Omega' + \Omega) \,+\, \delta(\omega - \Omega + \Omega')\right) \quad (2.31)$$

For the model of independent particles, for which the $\varrho_k(\omega)$ effectively reduces to the delta function $2\pi\delta(\omega - e_k)$, the correlation function becomes

$$\psi''_{\text{ipm}}(\omega) = \pi \sum_{jk} |\,F_{jk}\,|^2 \, n(e_j)(1 - n(e_k))\Big(\delta(\omega - (e_k - e_j)) \,+\, \delta(\omega + (e_k - e_j))\Big) \quad (2.32)$$

Comparing to (29) it is easy to deduce that in this limit the $\psi^0$ becomes:

$$T\left(\chi^{\text{T}} - \chi(0)\right)_{\text{ipm}} = \psi^0_{\text{ipm}} = \sum_{\substack{j,k \\ e_j = e_k}} |F_{jk}|^2 n(e_k)(1 - n(e_k)) = T \sum_k \left|\frac{\partial n(e)}{\partial e}\right|_{e=e_k} \left(\frac{\partial e_k}{\partial Q}\right)^2 \quad (2.33)$$

The last equation follows because of the two relations

$$-T\frac{\partial n(e)}{\partial e} = n(e)(1 - n(e)) \quad (2.34)$$

and

$$\hat{h}(Q) \mid k(Q)\rangle = e_k(Q) \mid k(Q)\rangle \quad \longrightarrow \quad F_{jk}\,\big|_{e_j = e_k} = \delta_{jk}\frac{\partial e_k}{\partial Q} \quad (2.35)$$

Here, the first one is a simple consequence of the form of the Fermi function. The second property is correct provided the scalar product $\langle j(Q) \mid \partial/\partial Q k(Q)\rangle$ remains finite for $e_j \to e_k$.

Let us turn to numerical estimates now. They have been obtained from calculations for quadrupole vibrations around a sphere within some schematic models. Details will be presented in the appendix *sect.*{A.1}. For the present purpose we take as single particle model the one of the infinitely deep square well. The equilibrium deformation is fixed to be a sphere for all temperatures. For such a case the difference between isothermal and adiabatic susceptibilities vanishes, a feature which can easily be understood on the basis of eq.(25), either by looking at the free energy or, if necessary, by direct evaluation with the help of the matrix elements of $\hat{F}$. These facts simplify the discussion on ergodicity, as we may fully concentrate on the evaluation of $\chi^{\text{T}} - \chi(0)$.

In Fig.2.2 we show the $\psi''(\omega)$ (top), together with the dissipative part of the response function $\chi''(\omega)$ (bottom), for $T = 1$ MeV on the left side, and for $T = 2$ MeV on the right side. First of all, we observe that, because of "collisional damping", the heat pole in (29), which we may identify as $\psi^0 2\pi\delta(\omega) \equiv {}_0\psi''(\omega)$, now has acquired a finite width, denoted by $\Gamma_T$ in the sequel. It is natural to approximate the functional form by a Lorentzian. In this sense we may write

$$_0\psi''(\omega) = \psi^0 2\pi\delta(\omega) \qquad \Longrightarrow \qquad {}_0\psi''(\omega) = \psi^0 \frac{\hbar \Gamma_T}{\hbar^2 \omega^2 + \Gamma_T^2/4} \qquad (2.36)$$

This function is normalized such that an integration of over $\omega$ gives back the $\psi^0$. Both $\psi^0$ as well as $\Gamma_T$ may be obtained by just fitting the Lorentzian (36) to the actual peak as it comes out from the computed correlation function, like shown in Fig.2.2. The $\psi^0$ can easily be deduced from ${}_0\psi''(\omega)$ by way of $\psi^0 = (\hbar \Gamma_T / 4) {}_0\psi''(\omega = 0)$. According to (30) the residue is related to the average of the squared zero-frequency part of our operator $\hat{F}$. Within our model, we should thus calculate ${}_0\psi''(\omega)$ from those terms in $\psi''(\omega = 0)$ for which $e_k = e_j$. Using for $\psi''_{jk}(\omega)$ the form given in (31) one finds

$$\begin{aligned}
{}_0\psi''(\omega = 0) &= \sum_{\substack{j,k \\ e_j = e_k}} |F_{jk}|^2 \psi''_{j=k}(0) \\
&= \sum_{\substack{j,k \\ e_j = e_k}} |F_{jk}|^2 \int_{-\infty}^{\infty} \frac{d\Omega}{2\pi} \, n(\hbar\Omega) \left(1 - n(\hbar\Omega)\right) \varrho_k(\Omega) \varrho_k(\Omega)
\end{aligned} \qquad (2.37)$$

Please notice that for the second factor the restriction $e_j = e_k$ would imply $k = j$, even for a general single particle spectrum without (35) being given. This is due to the special choice made in (15) and (16) for $\varrho_k$.

As a result from such a fit we present in Fig.2.3 by the solid line $\Gamma_T$ as function of $T$. The dashed curve represents twice the single particle width $\Gamma(\hbar\Omega = \mu, T)$ calculated from (16), but with the frequency fixed at $\hbar\Omega = \mu$. Both curves are practically identical, with slight deviations occurring only at large values of temperatures. This can be traced back to the fact that it is only the behavior at small frequencies $\omega$ which matters for the heat pole. It is interesting to see that for quite a large range of intermediate temperatures the $T$-dependence of $\Gamma_T$ turns out linear, following the simple rule $\Gamma_T \approx 2\Gamma(\mu, T) \approx 2T$. Finally, we should like to mention that $\Gamma_T$ reaches quite large values. This may perhaps come from the fact that by using a canonical distribution the importance of the heat pole is grossly overestimated.

In Fig.2.4. we show $\psi^0 / T = \chi^T - \chi(0)$ as function of temperature. The solid curve corresponds to collisional damping, the dashed one to independent particle motion as given by (33). It is observed, that a) in both cases the difference tends to zero for $T \to 0$, and that

b) it takes on finite values at $T \neq 0$, which increase with $T$ to level off above $T \approx 2$ MeV, and that c) the curve for collisional damping essentially is identical to the independent particle model.

The first result reflects an exact relation, telling us that in this sense both computations are in accord with it. The fact that for increasing $T$ the $\psi^0/T = \chi^{\mathrm{T}} - \chi(0)$, which for the present model is identical to $\chi^{\mathrm{ad}} - \chi(0)$, becomes different from zero tells us that the system studied is *not* ergodic. For the *independent particle model* this result is *not surprising*. The level scheme of the underlying single particle model is of *"diabatic"* nature rather than to reflect *"adiabatic"* behavior. Without considering an interaction, which would repel the levels, the spectrum has many degeneracies, such that the first one of our conditions is not fulfilled. The third point c) tells us that the way we treat the residual interaction $\hat{V}_{\mathrm{res}}^{(2)}(\hat{x}_i, \hat{p}_i)$ by collisional damping does not reduce the strength of the heat pole as one should expect it for an ergodic system. Various reasons may conceivably be responsible for this feature. We try to simulate effects of $\hat{V}_{\mathrm{res}}^{(2)}(\hat{x}_i, \hat{p}_i)$ by introducing complex self-energies. It could be that this approximation is too drastic to describe correctly the influence of $\hat{V}_{\mathrm{res}}^{(2)}(\hat{x}_i, \hat{p}_i)$ at very low frequencies. But even on this level improvements may be envisaged. As discussed at the end of *sect.*{2.2} the form (16) chosen for the imaginary part actually does not lift the degeneracies. Rather than modifying this form (16) for the self-energies, which has proven convenient for numerical applications, we suggest to "cure" the problem in a different, more pragmatic way.

Based on the observation that a correct treatment of the full residual interaction would imply level repulsion, we may just argue that the system in equilibrium can be expected to show ergodicity. Thus to get (7) to be fulfilled, we simply "prune" or "trim" the strength $\psi^0$ of the heat pole from the value it has in the *diabatic* limit, the *pure* independent particle picture, to the one expected for a decent thermal equilibrium. This means to enforce

$$\frac{1}{T}\psi^0 = \chi^{\mathrm{T}} - \chi(0) = \left(\chi^{\mathrm{T}} - \chi^{\mathrm{ad}}\right) - \left(\chi^{\mathrm{ad}} - \chi(0)\right) \qquad \longrightarrow \qquad \frac{1}{T}\psi^0 = \chi^{\mathrm{T}} - \chi^{\mathrm{ad}} \quad (2.38)$$

To justify this modification we have invoked the assumption that our system be close to thermal equilibrium. Such a situation can be found, for instance, when the system cannot move away quickly, as may be given for nuclear fission, in distinction to the entrance phase of a heavy ion collision. Then the system has enough time to explore the "adiabatic" landscape, in the sense that all kind of residual interactions may come into play. Since the model underlying the computation fails to reproduce such features we may suppose to simulate them by requiring (38). In the next subsection we will exhibit implications for the friction coefficient. This discussion will be completed in *chap.*{5} where we will contrast results of numerical computations done with and without considering (38).

Please notice, that by requiring (38) we also account for condition ii) found necessary to have ergodicity: the distribution in the total energy must be sufficiently narrow. We may recall from the discussion given at the end of part a) of this subsection, that the width given by the canonical distribution may not be small enough. Since in the nuclear case such a distribution is taken for convenience only, we need not necessarily take over its main deficiency when calculating a quantity which is particularly sensitive to energy fluctuations.

For the present case of studying vibrations around an equilibrium whose shape stays spherical for all $T$ the requirement (38) can be fulfilled in a particularly simple fashion. We know that in such a case the difference $\chi^{\mathrm{T}} - \chi^{\mathrm{ad}}$ vanishes identically. Therefore, by inspection of (38) and (35) we realize that we simply have to leave out all contributions from *diagonal* matrix elements $F_{j=k}$.

The result such a restriction has on the dissipative part of the response function is shown in the bottom part of Fig.2.2. Two computations are presented: the fully drawn line is obtained when in (24) all possible matrix elements $F_{jk}$ are included, for the dashed one the diagonal elements are excluded. It has been checked numerically, that both cases are accord with the fluctuation dissipation theorem, which for the present purpose we like to write in the version

$$\chi''(\omega) = \frac{1}{\hbar} \tanh\left(\frac{\hbar\omega}{2T}\right) \psi''(\omega) \tag{2.39}$$

### 2.6.2 Consequences for friction

In the previous subsection we have tried to argue about the size of the heat pole, or more precisely of its strength. Now we want to examine the consequences it would have on the friction coefficient. In this way we will understand its importance in the "diabatic" picture.

The heat pole involves the behavior of the intrinsic system at small frequencies. As for friction it will thus mainly be the coefficient in zero frequency limit which is effected. The general expression for this coefficient is given by (23), with eq.(24) representing the form derived for collisional damping. The following discussion will have a lot to do with the question of how physically one should interpret the limit $\omega \to 0$.

The dissipation fluctuation theorem (39) facilitates to express the contribution of the heat pole to the zero frequency limit of friction, called $_0\gamma(0)$ below. Indeed, differentiating (39) once with respect to $\omega$ and putting $\omega = 0$ afterwards one gets

$$\gamma(0) = \left.\frac{\partial \chi''(\omega)}{\partial \omega}\right|_{\omega=0} = \frac{\psi''(\omega=0)}{2T} \tag{2.40}$$

By use of (29) together with (36) and (30) we obtain

$$_0\gamma(0) = \frac{4\hbar}{\Gamma_T} \frac{\psi^0}{2T} = \frac{2\hbar}{\Gamma_T}\left(\chi^{\mathrm{T}} - \chi(0)\right) \tag{2.41}$$

We recall from Fig.2.4 that the factor $\chi^{\mathrm{T}} - \chi(0)$ comes out almost the same no matter whether or not we include collisional damping in the sense of eq.(16). The pre-factor $2/\Gamma_T$ can be justified only within collisional damping, of course. Fig.2.3 teaches us that this factor may be estimated to high accuracy from the single particle width by putting $\Gamma_T = 2\Gamma(\mu, T)$ with the $\Gamma(\omega, T)$ being given by (16). Estimating $\chi^{\mathrm{T}} - \chi(0)$ by its form (33) valid in the independent particle model we get for this component of friction

$$_0\gamma(0) = \frac{\hbar}{\Gamma(\mu, T)} \sum_k \left| \frac{\partial n(e)}{\partial e} \right|_{e=e_k} \left( \frac{\partial e_k}{\partial Q} \right)^2 \qquad (2.42)$$

From this form, together with the content of Fig.2.4 we may deduce the temperature dependence of $_0\gamma(0)$. Let us look first at large $T$, for which the $\chi^{\mathrm{T}} - \chi(0)$ approaches a constant value, implying that the $T$-dependence of $_0\gamma(0)$ is governed entirely by the one of $\Gamma(\mu, T)$.

(i) In case that the parameter $c$ of (16) is chosen to be *finite*, the $_0\gamma(0)$ approaches a finite value being proportional to $c^2/\Gamma_0$. As a peculiar feature, this limiting value of $_0\gamma(0)$ happens to be close to the value of the wall formula if for $c$ and $\Gamma_0$ the "standard choice" is used: $\Gamma_0 = 33.3$ MeV and $c = 20$ MeV.

(ii) Conversely, for $1/c = 0$, the case with which commonly relaxation times are estimated (cf.eq.(3.5) below), the $_0\gamma(0)$ would tend to zero like $T^{-2}$ and in this sense show a behavior typical of "hydrodynamical dissipation" or "two-body viscosity".

For small $T$ the pre-factor $1/\Gamma(\mu, T)$ behaves like $T^{-2}$ in both cases. But this divergence is cancelled by the term $\chi^{\mathrm{T}} - \chi(0)$, which is expected to approach zero exponentially, such that $_0\gamma(0)$ starts from zero and increases quickly to some maximum value at $T \approx 1$ MeV. For a graphical demonstration see Fig.2.5. At intermediate temperatures, the $_0\gamma(0)$, as given by (41) or (42), takes on very large values. This feature is attributed to the fact that perhaps our model grossly overestimates the magnitude of $\chi^{\mathrm{T}} - \chi(0)$. Indeed, provided the nucleonic degrees of freedom would behave ergodic, in the sense of eq.(7), the $_0\gamma(0)$ would become proportional to $\chi^{\mathrm{T}} - \chi^{\mathrm{ad}}$ (mind (38)). Also this difference is governed by diagonal matrix elements, but for the nuclear case it turns out much smaller than $\chi^{\mathrm{T}} - \chi(0)$. For temperatures above $T \approx 2$ MeV, where shell effects supposedly disappear, the mixed derivative of the free energy with respect to $Q$ and $T$ will become small such that $\chi^{\mathrm{T}} - \chi^{\mathrm{ad}}$ will approach zero (cf. (25)). As mentioned before, for the present numerical model the $\chi^{\mathrm{T}} - \chi^{\mathrm{ad}}$ even vanishes identically.

So far we have not been looking at the contributions to dissipation which come from the remaining part $_R\psi''(\omega)$ of the correlation function, or the corresponding one in the response function. Following our previous discussion, these contributions can be classified as those coming from those matrix elements $F_{jk}$ where the two energies are different $e_j \neq e_k$. In the lower part of Fig.2.2 the dashed curve corresponds to the situation in which for the present model the heat pole contribution has been removed. This figure is very instructive. First of all, it shows very clearly into which regime of frequencies this (perhaps fake) heat pole "scatters". Secondly, and more important, it demonstrates very clearly how this large contribution to friction comes about: It is this very tiny bump in the strength distribution at very small frequencies, which because of its large slope contributes so much to friction in the zero frequency limit.

Indeed, looking at the lower part of Fig.2.2 one may be inclined to define an average slope of the response function by smoothing over a range of a few MeV. As can be seen from this figure, such a slope would be close to the one of the dashed curves (in the lower parts). Notice that it is the response of the *intrinsic or nucleonic* degrees of freedom we are talking about, whose main excitation happens to be at larger frequencies. If we were to parameterize the latter in terms of a Lorentzian, for instance to simulate the first big peak in the strength distribution, the small wriggles at small frequencies would be washed out, indeed. The friction coefficient of the zero frequency limit which could be attributed to such a reduced slope would then be much smaller than the one given by (42) (and shown in Fig.2.5 by the fully drawn curve). Its value would be close to the one obtained by applying (40) to the $_R\psi''(\omega)$ discussed above. The result of such a computation is shown in Fig.2.5 by the curve being marked by squares. It shows the influence of ergodicity to friction, a point on which we will elaborate further later in the text. We should like to mention that such a picture has been adopted also in the computations discussed in *sect.*{2.5}.

## 2.7 Provisional stock-taking

Before closing this chapter we want to give a short summary of the results found about dissipation and add a few comments.

1) In general, we may isolate two contributions, one coming from the heat pole, the other one from the remaining part of the strength distribution.
2) For an ergodic system, the contribution from the heat pole becomes proportional to $\chi^{\mathrm{T}} - \chi^{\mathrm{ad}}$, and thus can be expected to be small and to decrease quickly with increasing $T$.
3) The situation is then similar to the case of hydrodynamics (see e.g. Chap.30 of [22]). There the attenuation of density waves gets two contributions, with the one from the heat pole (also called Landau-Placzek peak) being proportional to the difference between the isothermal and isentropic compressibility (which in turn is proportional to the difference between the specific heats at constant pressure and constant volume).
4) According to the previous discussion, the notion ergodic is very much related to what in nuclear physics often has been associated to the notion "adiabatic". On one hand, one may say that ergodicity requires non-degenerate spectra, a situation which is obtained easier for an "adiabatic" level scheme showing level repulsion. As for dynamics, the concept "adiabatic motion" implies that the system follows the lowest possible configurations with respect to the static energy. These configurations ought to be interpreted in the sense of the compound nucleus, and the static energy should be understood as the internal energy at given entropy.

5) Conversely, if the system behaves non-ergodic, or "diabatically", the contribution from the heat pole will be proportional to $\chi^T - \chi(0)$, as given within the single particle model. Since this quantity then will be large, the heat pole contributes sizable to friction.

6) Anticipating results from the next chapter, it can be said that the feature just mentioned is in accord with findings of two earlier papers, [15] and [17].

## 3 Linearized "Dissipative Diabatic Dynamics"

The strength function found in [14] for vibrations about equilibrium at larger temperatures exhibits features of macroscopic motion, not seen for the diabatic case shown in Fig.2.1. Whereas the latter is largely governed by shell effects their influence apparently disappears for the "adiabatic case" above some critical value of T. It should be of interest to see what happens in a theory like the one for "dissipative diabatic dynamics" (DDD) of Nörenberg et.al. (see e.g.[16]) which incorporates both "diabatic" and "adiabatic" features. In the next section we will apply the basic EOM of [16] to discuss vibrations in the language of response theory. As we will see, for large $T$ these vibrations of DDD show features typical of hydrodynamics, not only for the inertia, but for the friction coefficient as well.

Let us take eqs.(2.30-32) of [16], write them in first order in $q - q_0$ and add a term $-q_{\text{ext}}(t)$ on the right hand side. The latter represents an external force (with coupling $\delta H = q_{\text{ext}}(t)\, q$ to the system). For the sake of simplicity we assume the intrinsic relaxation time $\tau_{\text{intr}}$ to be constant. This implies to write for the non-Markovian force:

$$\int_{t_0}^t ds\ K(t,s)\dot q(s) = C_D \int_{t_0}^t ds\ \exp\left(-\frac{(t-s)}{\tau_{\text{intr}}}\right)\dot q(s) \qquad (3.1)$$

To facilitate direct comparison with our standard linear response treatment, we want to apply Fourier transforms in ordinary fashion. This means to put the initial time $t_0$ equal to $-\infty$. Defining the response function for collective motion in the usual way (see above) one easily derives the following expression:

$$\chi^D_{\text{coll}}(\omega) = \frac{1}{-\omega^2 B + C_D \chi^D_q(\omega)(-i\omega) + C(0)} \qquad (3.2)$$

Here, $B$ is the inertia for irrotational flow and $C_D$ represents the difference between the stiffnesses of the diabatic and the adiabatic potentials $C_{\text{di}}$ and $C(0)$, respectively:

$$C_D = C_{\text{di}} - C(0) \equiv \frac{\partial^2 E_{\text{di}}}{\partial q^2}\big|_{q_0,S} - \frac{\partial^2 E_{\text{qs}}}{\partial q^2}\big|_{q_0,S} = \frac{1}{k} - \frac{1}{k_{\text{di}}} \qquad (3.3)$$

(On the right we show once more the relation (2.19) to the corresponding coupling constants). Furthermore, in (2) there appears the equivalent to what we call "nucleonic" response function, namely

$$\chi_q^D(\omega) \equiv \int_{-\infty}^{\infty} d(t-s) \, \Theta(t-s) \, \exp\left((i\omega - \frac{1}{\tau_{\text{intr}}})(t-s)\right) = \frac{i}{\omega + \frac{i}{\tau_{\text{intr}}}} \qquad (3.4)$$

Please notice its simple structure which just represents one single, but overdamped intrinsic mode. In conjunction with the discussion of the last chapter, this feature will help us below to understand the physical nature of the dissipation mechanism of DDD. In (4) the $\Theta$ function has been introduced to take care of the upper integration limit in (1); it renders the $\chi_q^D(\omega)$ to be a causal response function. For a situation close to thermal equilibrium the $\tau_{\text{intr}}$ can be estimated as

$$\frac{\tau_{\text{intr}}}{\hbar} = \frac{2 \times 10^{-22}\text{sec}}{T^2 \times 10^{-1}\hbar} = \frac{3}{T^2 \text{ MeV}} = \frac{30}{\pi^2 T^2 \text{ MeV}} \qquad [(T) = \text{ MeV}] \qquad (3.5)$$

(if one assumes that the total excitation energy is shared equally among all the particles with the energy per particle given by $T^2/10$). Please recall from the discussion in $sect.\{2.2\}$ that the value of $\tau_{\text{intr}}$ agrees with the $\Gamma$ of (2.16) if one identifies $\Gamma = \hbar/\tau_{\text{intr}}$, puts $c = \infty$ and neglects the frequency dependence, i.e. assumes the excitations to occur at the Fermi surface. After some calculation one ends up with the following expression:

$$\left(-\chi_{\text{coll}}^D(\omega)\right)^{-1} = \omega^2 B - C(0) - \frac{C_D}{\omega^2 \tau_{\text{intr}}^2 + 1}\left(\omega^2 \tau_{\text{intr}}^2 - i\omega \tau_{\text{intr}}\right) \qquad (3.6)$$

Let us present a numerical calculation of the dissipative (imaginary) part of this response function. Its frequency dependence allows to deduce directly the transition we want to study, namely the one above which essentially no strength is seen any more in the high frequency mode (recall the discussion in $sect.\{2.4\}$. To this end let us fix the adiabatic stiffness to $C(0) = B\omega_0^2$ with $\hbar\omega_0 = 1$ MeV. By writing $C_D = fC(0)$ the inertia $B$ scales out. Its value can be related directly to the total strength of the energy weighted sum, which is of no importance here. The strength distribution is readily evaluated from (6). We show it in Fig.3.1 for various temperatures (see also [38]). In this calculation $f = 100$ was put such that at $T = 0$ the giant resonance lies at about 10 MeV. The value of $f$ agrees with the one we found analyzing Strutinsky computations of static energies. The figure exhibits similarities to the results discussed before. It clearly demonstrates the existence of the transition we spoke of before. A very similar behavior is found also in [18] where a Landau-Vlasov approach is used. However, in both cases the transition appears at temperatures not smaller than 4 MeV, which is to say at values which are definitely larger than the ones found or suggested in our linear response model (see [14]). But there are other differences. It is seen that at small $T$ in this model (different to the one of [18]) there exist *no low frequency mode at all*—which certainly goes back to the scaling assumption made from the start. More important are differences in the transport coefficients which we are going to address now.

Inspecting the denominator (6) of the collective response function, its limiting forms are readily derived:

i) For $\omega\tau_{\text{intr}} \gg 1$ the effective stiffness is given by $C_D$ (being $\gg C(O)$), and the friction coefficient becomes proportional to $\tau_{\text{intr}}^{-1}$—like it is for zero-sound modes.

ii) For $\omega\tau_{\text{intr}} \ll 1$ the stiffness reduces to $C(0)$ and the friction coefficient becomes

$$\gamma_D = \tau_{\text{intr}} C_D \tag{3.7}$$

It is worth noticing that the inertia stays the same independent of frequency or temperature. In the following we want to discard the case (i) as it is only the second one which bears features of the "zero-frequency limit". For a slow mode like fission the transition temperature, valid for the present model, can be estimated to $T \approx 3$ MeV. This follows from (5) and by taking for the $\hbar\omega$ a value of the order of 2 MeV, as may be expected from the formula $\sqrt{C(0)/B}$ (c.f.[18]).

As for (ii) the stiffness turns out to be given by the "adiabatic" potential landscape, there is some relation to what we have called "adiabatic" motion. However, the friction force found here definitely is the one which we like to associate to the heat pole, but calculated in the "diabatic" picture. Indeed, the form (7) can be seen to be just given by (2.42), if we only identify $\tau_{\text{intr}}/\hbar = \Gamma^{-1}(\mu, T)$. This follows because within the independent particle model the $C_D$ can easily be seen to be given by $\chi^T - \chi(0)$; for details of such a proof we want to refer to [20]. Actually, in [15] rather than (7) the form (2.42) has been specified in eq.(5.12) to determine friction, with $\tau_{\text{intr}}$ being called $\tau_{\text{loc}}$. As a matter of fact, the association to the "heat pole" can directly be seen from the response function (4). It has just one pole lying at $\omega = 0$ whose width is given by $\Gamma_T/2 = \Gamma(\mu, T) = \hbar/\tau_{\text{intr}}$.

Finally, let us comment once more on the nature of such a friction coefficient being associated to the "heat pole". Being proportional to $\tau_{\text{intr}} \sim T^{-2}$ it shows similarities to the one of hydrodynamics. However, this form does not come from "collision dominance", and hence, has not much to do with the two-body viscosity of an ordinary liquid. We shall come back to this point in *sect.*{5.3}.

## 4 Strutinsky smoothing and collective dynamics in the independent particle model

In the previous chapters we have found evidence that strength distributions of isoscalar modes show a tendency to exhibit macroscopic behavior when temperature is raised. When calculating the total static energy, it has been proven successful to relate the notion of the macroscopic limit to averages over single particle degrees of freedom. We are now going to ask the question to which extent this concept can be taken over when looking at dynamical properties. Particular emphasis shall finally be laid on dissipation, the macroscopic limit of which has been suggested (for a recent review see [6]) to be represented by the wall formula [5].

Let us recall the basic points of the Strutinsky procedure: i) One starts from a description of independent particle motion, ii) averages out the "shell effects" and iii) claims that the averaged quantity represents the "true" macroscopic limit. As the latter is not supposed to be accessible theoretically, one is forced to deduce it by applying adequate procedures to experimental results. It is well known that for the classic application of the Strutinsky method the macroscopic limit can be defined at least in threefold fashion, a) by smoothing of the single particle spectrum, b) by increasing the thermal excitation of the system, and c) by smoothing over particle number (which is almost equivalent to performing the macroscopic limit in true sense, namely letting the nuclear size become large). For our purpose we will mainly be concerned with the first possibility, only to try to set up some loose relations to b). However, to establish connection to the wall formula, it will prove essential to stick to the picture of independent particles. This already rules out to apply concepts borrowed from the picture of the compound nucleus, on which we dwelled upon in *sect.*{2.6}.

We all know how perfectly well the Strutinsky procedure works for the case of the static energy. But there one is in the lucky situation that an average over mass number specifies the macroscopic limit almost by definition. For the dynamic case we do not even know whether or not a well defined macroscopic limit can be expected to exist. We have seen above that for nuclear dissipation various approaches lead to different results. Below we shall exploit this example further. But first we like to address the question of extending smoothing procedures to treat dynamical aspects in general.

## 4.1 Averaged response functions for nucleonic motion

First we would like to compare a few possibilities of averaging response functions, developing the suggestions presented in [39]. We will apply the strategy to smooth the dissipative part and to exploit Kramers-Kronig relations for obtaining the other ones, if needed. Within the independent particle model the dissipative response function can be written as

$$\chi''(\omega) = -\pi \sum_{jk} (n(e_k) - n(e_j)) \mid F_{jk} \mid^2 \int_{-\infty}^{\infty} de\, \delta(\hbar\omega - e + e_j)\delta(e - e_k) \qquad (4.1)$$

This follows without difficulties from (2.13) and (2.14) for $\varrho_k(\omega) = 2\pi\, \delta(\hbar\omega - e_k)$.

Quite generally, the smoothing procedure itself may be defined in the following way. Suppose we are given some function $G(x)$ of $x$ which is to be averaged over an interval $\Delta x$. Then we may define the average $\overline{G}(x)$ by

$$\overline{G}(x) = \int_{-\infty}^{\infty} G(x') \frac{1}{\Delta x} f\left(\frac{x' - x}{\Delta x}\right) dx' \qquad (4.2)$$

The smoothing function $f((x' - x)/\Delta x)$ has some bell-like shape of width $\Delta x$, with its maximum lying at $x' = x$. In practice we may want to take a Lorentzian, a Gaussian, or the function used in Strutinsky's shell correction method. The latter is defined as

$$f(x) = \frac{e^{-x^2}}{\sqrt{\pi}} \sum_{n=0,2,...}^{M} \alpha_n H_n(x) \qquad (4.3)$$

with the $H_n(x)$ being Hermite polynomials and the coefficients $\alpha_n$ being given by the following recurrence relations

$$\alpha_{n+2} = \frac{-\alpha_n}{2n}, \qquad \alpha_0 = 1. \qquad (4.4)$$

The smoothing function (3) represents nothing else but the first $M$ terms of an expansion of the $\delta$-function into series of Hermite polynomials. The characteristic feature of (3) and (4) is that this particular smoothing function restores any polynomial in $x$ which is of order $k \leq M$,

$$P_k(x) = \int_{-\infty}^{\infty} dy\, f(x - y)\, P_k(y), \qquad \text{if} \qquad k \leq M \qquad (4.5)$$

with $M$ being the highest order of polynomials considered in (3).

The most direct way is to smooth the response function over frequency itself, which is to say to use (2) with $x$ being $\omega$. Indeed, for a supposedly discrete single particle spectrum the $\chi''(\omega)$ will be a strongly oscillating function of $\omega$. Performing in (1) the integral over $de$ we get

$$\overline{\chi}''(\omega) = -\pi \sum_{jk} \left(n(e_k) - n(e_j)\right) \left| F_{jk} \right|^2 \frac{1}{\gamma_{\rm av}} f\left(\frac{\hbar\omega - (e_k - e_j)}{\gamma_{\rm av}}\right) \tag{4.6}$$

with the averaging interval in frequency being expressed in energy units, $\hbar\Delta\omega \equiv \gamma_{\rm av}$. One of the main reasons why (1) shows strong oscillations in $\omega$ is due to the presence of the $\delta$-functions. Realizing that the latter represent the density of single particle states, another procedure is suggested: Besides just smoothing over the frequency itself, we may as well average over the density. Such procedure was used in [39] and lead to the smooth response function of the type

$$\overline{\chi}''(\omega) = -\pi \sum_{jk} \left(n(e_k) - n(e_j)\right) \left| F_{jk} \right|^2 \frac{1}{\gamma_{\rm av}} f_2\left(\frac{\hbar\omega - (e_k - e_j)}{\gamma_{\rm av}}\right) \tag{4.7}$$

where

$$f_2(x) = \int_{-\infty}^{\infty} dy\, f(x-y)\, f(y) \tag{4.8}$$

Later on, this expression will be referred to as density smoothed response function, to be distinguished from the frequency smoothed expression (6) given above. Comparing (6) with (7) it is easily recognized from (8) that the second version can actually be obtained by averaging the first one a second time. The result for $f_2(x)$ then depends on the smoothing function $f(x)$ one starts with. For a Lorentzian the width just doubles and the density averaged response function is smoother than the one obtained from frequency averaging. If on the other hand, we take the smoothing functions to be those of the Strutinsky method proper, namely as given by (3), density averaging and frequency smoothing are about the same. This is due to the property (5) which guaranties that a repetition of averaging restores averaged quantities.

The two procedures just mentioned are somewhat unsatisfactory as in the basic expression (1) besides the $\delta$-functions there are other quantities which depend on the single particle states or their energies and which thus are amenable to fluctuations in the spectrum, namely the matrix elements $\left| F_{kj} \right|^2$. To account for this feature one needs to be able to smooth also over the spectral distribution.

To this end we begin replacing in (1)

$$\sum_{jk} \to \int de\, \overline{g}(e) \int de'\, \overline{g}(e') \tag{4.9}$$

where the original level density $g(e) = \sum_k \delta(e - e_k)$ has been substituted by the smooth one

$$\overline{g}(e) = \frac{1}{\gamma_{\rm av}} \int f\left(\frac{e-e'}{\gamma_{\rm av}}\right) g(e') de' = \frac{1}{\gamma_{\rm av}} \sum_k f\left(\frac{e-e_k}{\gamma_{\rm av}}\right) \tag{4.10}$$

Having the sums over states changed into integrals over energies requires to introduce averaged squared matrix elements. They may be defined in the following way

$$\mathcal{F}^2(e,e') = \sum_{jk} \left| F_{jk} \right|^2 f(\frac{e-e_k}{\gamma_{\text{av}}})f(\frac{e'-e_j}{\gamma_{\text{av}}}) / \sum_{jk} f(\frac{e-e_k}{\gamma_{\text{av}}})f(\frac{e'-e_j}{\gamma_{\text{av}}}) \qquad (4.11)$$

Replacing then the matrix elements $\left| F_{kj} \right|^2$ in (1) by their smooth counterparts (11) one will get for the averaged response function

$$\overline{\chi}''(\omega) = \pi \int de \left( n(e - \frac{\hbar\omega}{2}) - n(e + \frac{\hbar\omega}{2}) \right) \overline{g}(e - \frac{\hbar\omega}{2})\overline{g}(e + \frac{\hbar\omega}{2})\mathcal{F}^2(e - \frac{\hbar\omega}{2}, e + \frac{\hbar\omega}{2}) \quad (4.12)$$

To distinguish between (12) and (7) we will call (12) the *spectrally smoothed response function*.

Before closing this section we would like to add two remarks. The first one concerns the behavior of the averaged response with the averaging parameter $\gamma_{\text{av}}$. Trivially, for $\gamma_{\text{av}} \to 0$ the averaged functions turn into the original expression (1) no matter which procedure is employed. More interesting is the case of growing $\gamma_{\text{av}}$. Then we would expect quantum effects to gradually disappear such that for sufficiently large $\gamma_{\text{av}}$ one may expect to reach the *macroscopic limit*. The second remark concerns the choice of the smoothing function $f(x)$. As mentioned before, one could think of various options. From a pragmatic point of view, the simplest ones are offered by Lorentzians and Gaussians, and they may indeed serve to obtain valid guidelines. However, if it comes to question of stability etc., the smoothing function of the original shell correction method by Strutinsky [40] becomes advantageous.

### 4.2 Friction from averaged response functions

For friction the existence of a macroscopic limit in literal sense has been looked at in [41] applying a slab model to surface modes. Inside the slab of (average) width $L$ there was a gas of free nucleons. Their motion was treated with the Vlasov equation with boundary conditions simulating mirror reflections at the walls—*but discarding collisions among the particles*. This situation corresponds to the one of the wall formula. Indeed, in [41] the latter could be reproduced in the limit of $L \to \infty$. This should be no surprise as all existing derivations of the wall formula have been done assuming an at least semi-infinite system. The more exciting issue of the paper was, however, that this limiting value could as well be obtained for a *finite* $L$ when a Strutinsky smoothing was applied (actually, in the terminology from above it was "frequency smoothing"). So there is a model where for a *dynamical* quantity like friction the existence of a macroscopic limit can clearly be seen and for which this limit is recovered by Strutinsky smoothing.

In the following we want to investigate properties of the friction coefficient, defined in terms of smoothed response functions, but applied to *finite systems*. The question of particular interest is whether or not, for increasing $\gamma_{\text{av}}$, the friction coefficient $\gamma(0)$ reaches some constant value. To evaluate $\gamma(0)$ for the different smoothing procedures we need to calculate (2.23) for the functions (6), (7) and (12). The results for frequency and density smoothing may be found in [39].

The case of spectral averaging is somewhat delicate as here a frequency dependence appears also in the occupation numbers. As a matter of fact, to calculate the derivative of (12) at $\omega = 0$ it is only this part whose derivative has to be taken. This is easily recognized after inserting (12) into (2.23). The result can then be written in a form like (2.24)

$$\gamma(0) = -\hbar\pi \int de \, \frac{\partial n(e)}{\partial e} \, \{\bar{g}^2(e)\mathcal{F}^2(e,e)\} \tag{4.13}$$

The function

$$-\frac{\partial n(e)}{\partial e} = \frac{1}{4T\cosh^2(\frac{e-e_F}{2T})} \tag{4.14}$$

is of bell-like shape having a width $T$ and being peaked at the Fermi energy $e_F$. Under the condition that the $\bar{g}(e)$ and $\mathcal{F}^2(e)$ are smooth around $e = e_F$ on the scale of $T$, the integral (13) can be calculated approximately by expanding $\bar{g}^2(e)\mathcal{F}^2(e)$ to low order in $e - e_F$. This involves moments of $\partial n(e)/\partial e$ which can be calculated in terms of Bernoulli numbers. In this way one gets to order 2

$$\gamma(0) \approx \left(1 + \frac{\pi^2 T^2}{6}\frac{d^2}{de_F^2}\right)\bar{\gamma} \tag{4.15}$$

with the leading term being given by

$$\bar{\gamma} = \hbar\pi\bar{g}^2(e_F)\mathcal{F}^2(e_F, e_F) \tag{4.16}$$

Using the definitions (10) and (11) it can be written as:

$$\bar{\gamma} = \hbar\pi \sum_{jk} |F_{jk}|^2 \frac{1}{\gamma_{\text{av}}^2} f(\frac{e_k - e_F}{\gamma_{\text{av}}}) f(\frac{e_j - e_F}{\gamma_{\text{av}}}) \tag{4.17}$$

Different to the case of frequency or density smoothing, here diagonal matrix elements contribute. But since this formula is derived within and to be applied to the pure independent particle model, there is no reason to subtract this contribution.

It is this form (17) which we want to exploit below, both for analytical analysis as well as for numerical computations. The latter shall be restricted to spheroidal isoscalar quadrupole vibrations around a spherical equilibrium. This implies the one body operator $\hat{F}$ to be given by a radial part times the spherical harmonic $Y_{20}(\cos\theta)$, (see (A.1)). For the single particle motion two models will be used, the infinitely deep square well and the oscillator, discarding spin-orbit forces in both cases. Details are given in Appendix A.

### 4.2.1 The infinitely deep square well

For this model the squared matrix elements $F^2_{nl,n'l'}$ become separable (see (A.7)). Therefore the friction coefficient (17) simplifies to

$$\bar{\gamma} = 4\pi\hbar\epsilon_F^2 \sum_l \bar{g}_l(\epsilon_F)(h_{ll}\bar{g}_l(\epsilon_F) + h_{ll+2}\bar{g}_{l+2}(\epsilon_F) + h_{ll-2}\bar{g}_{l-2}(\epsilon_F)) \qquad (4.18)$$

The $\bar{g}_l(\epsilon_F)$ represent the density of levels at *fixed orbital angular momentum* taken at the Fermi energy,

$$\bar{g}_l(\epsilon_F) = \sum_n \frac{1}{\gamma_{\text{av}}} f(\frac{\epsilon_{nl} - \epsilon_F}{\gamma_{\text{av}}}) \qquad (4.19)$$

In the next chapter we will present numerical computations of the expression (18) and we will examine the question of whether or not $\bar{\gamma}$ will reach a plateau if taken as function of $\bar{g}_l(\epsilon_F)$. Here we want continue with some analytical studies.

For sufficiently large $\gamma_{\text{av}}$ the density $\bar{g}_l(\epsilon_F)$ is expected to be a smooth function both of $\epsilon_F$ and $l$ and one may thus use the approximation

$$\bar{g}_{l\pm 1}(\epsilon) \approx \bar{g}_l(\epsilon) \pm \partial \bar{g}_l(\epsilon)/\partial l \qquad (4.20)$$

Due to the degeneracy of single-particle states with fixed $l$ the contribution to (18) comes mainly from the states with $l \gg 1$, for which one may expand $h_{ll'}$ in powers of $1/l$. Using this expansion and (20) one gets to the leading order in $1/l$ the following expression for the friction coefficient (18)

$$\bar{\gamma} = \hbar\epsilon_F^2 \sum_l (2l+1)\bar{g}_l^2(\epsilon_F) \qquad (4.21)$$

So for the square well potential the averaged friction coefficient $\bar{\gamma}$ is determined solely by the Fermi energy $\epsilon_F$ and the average density of levels $\bar{g}_l(\epsilon_F)$, calculated at $\epsilon_F$ for fixed orbital momentum. To estimate the latter one may use the Thomas-Fermi approximation [42] to get

$$\bar{g}_l(\epsilon_F) = \frac{1}{2\pi\epsilon_F}\sqrt{(x_F^2 - l^2)}\,\Theta(x_F^2 - l^2) \qquad (4.22)$$

Here we used the abbreviation

$$x_F \equiv k_F R_0 = \frac{\sqrt{2m\epsilon_F} R_0}{\hbar} \qquad (4.23)$$

Within this approximation, in (21) the sum over $l$ can be replaced by an integral. Inserting (22) one finally gets:

$$\bar{\gamma} = \frac{\hbar x_F^4}{8\pi^2} \qquad (4.24)$$

This result is identical to the dissipation rate for classical particles colliding with a moving potential well [6, 43], details of which are given in the Appendix *sect.*{A.2}.

With this result at our disposal, we may now discuss for the spherical shape the same situation as mentioned before for the slab. Indeed, the Thomas-Fermi approximation used for $\bar{g}_l(\epsilon_F)$ becomes exact in the limit that both $R_0$ as well as $l$ become very large. But the same limit may eventually be obtained by averaging over the single-particle levels of a finite nucleus, which is to say by calculating $\bar{g}_l$ from (19). Examples of such computations of $\bar{g}_l$ for various $l$'s are shown in Fig.4.1 as functions of an averaging interval $\Delta$, the latter being related to the $\gamma_{\text{av}}$ by

$$\gamma_{\text{av}} = \frac{2x_F \hbar^2}{2mR_0^2} \Delta \tag{4.25}$$

with the $x_F$ defined by (23). For the square well potential it is more convenient to average in wave number rather than in energy, simply because the density of states is a polynomial not in $e$ but in $k_F$. It is observed that the $\bar{g}_l$'s reach plateaus which starts at the somewhat large values of $\Delta \approx \pi$. This feature can be traced back to properties of the zeros of Bessel functions. Asymptotically, the latter behave like

$$j_l(x \to \infty) \to \frac{1}{x}\sin(x - l\pi/2) \tag{4.26}$$

implying that the zeros become equidistant with its spacing being equal to $\pi$. In Fig.4.1 we also plot, as the solid, dotted curve, the total density of levels as given by

$$\bar{g}(x_F) = \sum_l (2l+1)\bar{g}_l(x_F) \tag{4.27}$$

It has a perfect plateau which is reached at the smaller values of $\Delta \approx 1.0$. This value corresponds to $\gamma_{\text{av}} \approx 10 MeV$, which is to say that value at which the shell structure in the single-particle spectrum is washed out. The very fact that for the friction coefficient the plateau must be expected at the larger value of $\Delta \approx \pi$ has some implications. Firstly, it may cause problems for calculations using potentials of finite depth. The number of states with fixed orbital momentum may be too small to allow for the unambiguous definition of average quantities. Secondly, translating to temperature and assuming shell effects in usual sense to disappear at $T = 2MeV$, one would expect the plateau for friction coefficient appear at about $T \approx 4$ MeV.

Before closing this subsection we like to add a remark on the temperature dependence of the friction coefficient of the wall formula. Going back to (15) one may get the Sommerfeld expansion for the averaged friction (24). Using (23) one obtains:

$$\gamma(0) = \frac{\hbar x_F^4}{8\pi^2}\left(1 + \frac{1}{3}\pi^2\left(\frac{T}{e_F}\right)^2\right) \tag{4.28}$$

Such a form has been derived before in [43]. Later on we will come back to the question about the validity of this temperature dependence.

### 4.2.2 The oscillator potential

The expression (17) for the friction coefficient in the case of oscillator potential is somewhat more complicated as compared to the square well potential. Then the friction coefficient cannot be expressed in the form (21). On the other hand, because of selection rules for both the angular and the radial parts of $\hat{F}$, the double summation in (4.17) reduces to a single one. Results will be shown in *chap.*{5}.

## 5 Discussion of numerical results for friction

Above we have discussed various theoretical possibilities of describing collective motion. In this chapter we want to examine some implications the various models have on the friction coefficient $\gamma$. In particular we want to examine the following points or questions:
1) The macroscopic limit in the collisionless regime.
2) Why are collisions important?
3) How does $\gamma$ vary with $T$?
4) When does $\gamma$ show "hydrodynamical" behavior?
5) In which way does that depend on details of collisional damping?
6) The difference between ergodic and non-ergodic systems.

These items are put in arbitrary order. We will discuss them while presenting numerical results. All computations are still of somewhat schematic nature. Rather than being able to present final results we hope to deliver guidelines which might be worth considering in comparisons with experiments. Some preliminary conclusions, which last not least are based on experimental findings will be reserved to the next chapter.

### 5.1 Friction from averaging procedures

First we want to take up the problem of getting wall friction as the macroscopic limit for a model of independent particles. In *subsect.*{4.2.1} we have seen the wall formula to appear from the Thomas-Fermi model. There we have argued that in this model in some sense one still looks at large systems. We have speculated that eventually the same limit can be simulated by Strutinsky smoothing, if applied to a nuclear system of realistic size. Such a goal is reached if, as a function of the averaging parameter $\gamma_{\text{av}}$, the friction coefficient $\overline{\gamma}$ reaches a plateau.

In the case of square well potential the dependence of $\overline{\gamma}$ on $\gamma_{\text{av}}$ may be obtained from eq.(4.18) after calculating the density $\bar{g}_l(\epsilon_F)$ from (4.19) as the sum over levels having the same $l$. In this case the matrix elements have no natural cut-off, say as a function of the single particle energy $e_k$. We have thus limited the summations in formulas like (4.19) to energies $E_{\text{cut}} = 30 MeV$ or $E_{\text{cut}} = 100 MeV$ above the Fermi energy. The result of such computations are shown in the left hand side of Fig.5.1. The value of $\overline{\gamma}$ is seen to have a plateau. The quality of the latter depends on the cut-off parameter $E_{\text{cut}}$ as well as on the degree $M$ of the smoothing function (4.3). In the right hand side of Fig.5.1 a similar calculations is shown for the oscillator potential instead of the square well. In all cases the value for the wall formula (A.16) is shown by the line with stars as a reference value. For the case of the oscillator potential the $x_F$ was related to the smoothed Fermi energy

$$\frac{\hbar^2 x_F^2}{2mR_0^2} = \bar{\epsilon}_F = (N_F + 2)\hbar\omega_0 \qquad (5.1)$$

where $N_F$ is the number of occupied shells in spherical oscillator potential.

The fluctuations seen in the region of small $\Delta$, in the case of spectral smoothing, have no physical meaning. They are a consequence of using the smoothing function (4.3) which can be negative due to the presence of polynomial in (4.3). They would be absent, for example, if one were to apply a Gaussian smoothing function, ($M = 0$ in (4.3)), for which one would not find a plateau, of course.

One may speculate whether or not Strutinsky smoothing might be related to or could perhaps include in some way effects of collisional broadening. Recall please that the averaging interval is of the order of 10 MeV, a distance which certainly is larger than the average interval in energy over which nucleon's will effectively scatter even at large excitations. On one hand this could mean that the effects of scattering of nucleons are hidden in such an averaging procedure. On the other hand, it may as well be such that the smoothing is so crude as to wipe out any sensitivity of collisional broadening to details of the specific excitations of the system as a whole or of particular states.

To answer such questions let us look at some formal details, once more. It is not difficult to see that eq.(4.12) can in some sense be related to (2.14). There is of course the difference in that in (4.12) smoothed squared matrix elements have been used. But they could be introduced in (2.14) as well, which actually would not make a big difference provided the other factor, $\chi''_{jk}(\omega)$, is sufficiently smooth already. The latter will be so for sufficiently broad $\varrho_k$ and these quantities are easily related to the smoothing function $f(x)$ which appears in the $\bar{g}$ of (4.10). All what is needed is (i) to identify the $\Gamma/2$ of (2.15) by the *constant* $\gamma_{\text{av}}$ appearing through (4.10) in (4.12) and (ii) to put equal zero the real part of the self energy. However, doing so the *form* of the smoothing function is determined by (2.15), which then degenerates to an ordinary Lorentzian, *definitely not* to the form (4.3) used in the Strutinsky method proper.

The original picture underlying the wall formula is the one of independent particle motion. This is valid as long as the mean free path is much larger than the dimension of the system. Let us do some simple estimates to see whether or not such a situation is given. In nuclear matter the mean free path is given by $\lambda = 1/2k_I$, with $k_I$ being the imaginary part of the wave number $k$. The latter can approximately be related to the imaginary part of the self-energy. In principle, one ought to take into account the frequency dependence of $\Gamma$ and care about the difference between the k-mass and the nucleon mass (for a nice discussion we like to refer to [44]). But for our purpose we may just neglect such refinements and write

$$\lambda = \frac{1}{k_I} \approx \frac{k_R}{m}\frac{1}{\Gamma} = \frac{k_F}{m}\frac{1}{2\gamma_{\mathrm{av}}} \tag{5.2}$$

Here the real part of the wave number has been estimated by the Fermi momentum. The crucial quantity is the ratio $r = \lambda/R_0$ with $R_0$ being the nuclear radius. For a Fermi gas the latter can be related to the Fermi wave number. Putting everything together one gets

$$r \equiv \frac{\lambda}{R_0} \approx \frac{k_F}{m}\frac{1}{2\gamma_{\mathrm{av}}}\left(\frac{9\pi A}{4}\right)^{-\frac{1}{3}} \quad k_F = \frac{1}{2\Delta} \tag{5.3}$$

using the relation $\Delta = \gamma_{\mathrm{av}} m R_0/(\hbar^2 k_F)$ between the $\gamma_{\mathrm{av}}$ and the parameter $\Delta$ for the square well. For increasing $\Delta$ the mean free path becomes much smaller than the nuclear dimension. Looking at Fig.5.1 it is seen that a plateau is not reached, unless $\Delta$ becomes larger than one. But for the finite nucleus we look at, the wall formula can only be justified within the plateau. The conclusion one is inclined to draw from this fact is to say that collisions can hardly be discarded.

## 5.2 Friction and collisional damping

In this section we like to present results for the friction coefficient within linear response theory as outlined in *chap.*{2}. The response functions for internal motion are being calculated taking into account the effects collisional damping has on the particles' self-energies.

Some results have already been shown in the table in *sect.*{2.5}, for the case of quadrupole vibrations of lead. There all three transport coefficients have been considered. They have been obtained from an analysis of the collective response function (2.8) and thus account for self-consistency. For the present discussion we concentrate solely on the friction coefficient $\gamma$. As a very characteristic feature, it is observed that $\gamma$ *increases* with temperature, in clear distinction to the case of the wall formula or of "hydrodynamic models". This increase has nothing to do with the one seen in Fig.2.5 which we were able to attribute to the heat pole. As mentioned earlier, such a contribution actually has been left out in the computation presented in *sect.*{2.5}. It will be left out also for the computation to be presented now, for which we are going to concentrate on the influence of the self-energy. More precisely, we want to study the influence of the functional form given in (2.16). To this end we have performed calculations for friction in the zero frequency limit, using for the single particle motion the schematic models mentioned before; for details we may again refer to the appendix *sect.*{A.1}.

Let us turn to the numerical results obtained for our two schematic models. In Fig.5.2 we show the temperature dependence of friction. Let us first concentrate on the fully drawn curve which corresponds to an evaluation of (2.24), with the proportionality constant in (2.16) chosen to be $1/\Gamma_0 = 0.06\ MeV^{-1}$. At small temperature, we see a marked increase with $T$ which is typical for friction in the zero frequency limit, and which should thus be taken with reservations only, say below $T = 1 \cdots 2\ MeV$. At larger temperatures, we see $\gamma(0)$ to reach some maximal value. In the case of the square well potential it is of the same order of magnitude as the one of the wall formula, shown here by the horizontal line.

Comparing the upper and lower half of the Fig.5.2, we observe a pronounced sensitivity of the $T$-dependence of friction on the cut-off parameter $c$. For the case of having in (2.16) the $c \to \infty$, friction drops like $T^{-2}$, as expected for hydrodynamic modes. For the present value of $1/\Gamma_0 = 0.06\ MeV^{-1}$, the transition occurs at $T \approx 4 \cdots 5\ MeV$. For the smaller value of $1/\Gamma_0 = 0.03\ MeV^{-1}$ (cf.(2.17)) it would be at the somewhat higher value of $T \approx 6\ MeV$. Conversely, using a finite cut-off $c$, friction is found to level off at larger temperatures.

So far we have not commented on the second curve shown in Fig.5.2. It represents a computation in which the single particle strength $\varrho_k(\omega)$ is evaluated in a kind of "on-shell" approximation. By the latter we mean to replace in the form given by (2.15) the $\hbar\omega - \mu$ appearing in the self-energy by the $e_k - \mu$. Essentially that means to approximate the one body Green function to which the $\varrho_k(\omega)$ is associated to by the "on-shell" function, which sometimes is referred to as the quasi-particle approximation. We see that this approach overestimates dissipation, a feature which is in common with the result found in [26] for the width of giant resonances. The same seems to be true for friction at small to moderately high temperatures. This is another nice demonstration of the intriguing structure behind the forms (2.14)-(2.16), (2.24) with their delicate convolution integrals.

The results presented so far in this section were obtained by enforcing ergodicity to be given in the sense of (2.7). For the spherical models used this implies to neglect any contribution from the heat pole, as here even the difference between isothermal and adiabatic susceptibility vanishes. Conversely, it is clear that exactly for the sphere with its many symmetries the violation of (2.7) in a mere independent particle model will be particularly large. One should expect it to be much smaller for the deformed, more irregular shapes encountered in fission. It is reassuring that in the preliminary results of [31] the $T$-dependence of friction discussed in $sect.\{2.5\}$ have been confirmed, which were attained for deformations along a fission path, discarding the heat pole.

Nevertheless, for the spherical model taken here, let us show once more the contribution from the heat pole to friction. Different to the case presented in Fig.2.5, where *only* the heat pole component $_0\gamma(0)$ was considered, as given by (2.41), we are now going to take the full response function. The results for the square well are shown in Fig.5.3. They have been computed for two different cut-off parameters $E_{\text{cut}}$ for the single particle basis (see the appendix *subsect.{A.1.1}*). As seen from the figure this parameter does not play a crucial role here. Again details of the choice of $\Gamma(\hbar\omega, T)$ matter somewhat more. We show three cases, obtained for i) taking the full form as given by (2.16) (fully drawn line), ii) for the case short-dashed curve) that $\Gamma(\hbar\omega, T)$ is approximated by $\Gamma(\mu, T)$, the value which relates to the width of the heat pole, and finally iii) the on-shell approximation $\Gamma(\hbar\omega = e_k, T)$ (long-dashed curve) discussed before. For temperatures below $T \simeq 2\ldots 3$ MeV the fully drawn curve is close to the corresponding one of Fig.2.5. Moreover, for all $T$, the sum of the solid and the "solid-squared" curves in Fig.2.5 is very close to the fully drawn line in Fig.5.3. Like before, the on-shell approximation leads to results which differ considerably from the exact ones. The approximation of using a for $\Gamma(\hbar\omega, T)$ the frequency independent value $\Gamma(\mu, T)$ seems to be rather good. This is due to the fact that the zero-frequency limit of friction we look at here involves low frequency excitations only.

### 5.3 Hydrodynamic behavior

Often one speaks of hydrodynamic behavior if the friction coefficient changes with temperature like $\propto T^{-2}$. We believe this notion to be somewhat misleading as genuine two body viscosity can arise only in case that collective dynamics is dominated by collisions. Looking back at the results presented in Fig.5.2, one is inclined to attribute such a behavior at best to the regime of large temperatures, certainly not below $T \approx 3 - 4$ MeV. There indeed, $\gamma(0)$ is seen to be roughly $\propto T^{-2}$, with the precise behavior depending both on details of the shell model as well as on the choice of the particles' self-energies. A relation like $\gamma \propto T^{-2}$ is seen in cleanest form whenever the frequency dependence of $\Gamma(\hbar\omega, T)$ is neglected, be it by having $\Gamma(\hbar\omega = e_k, T)$ or by by using $\Gamma(\hbar\omega = \mu, T)$, like it is done in estimates of the relaxation time.

The very fact that in this regime such a behavior implies "collision dominance" may be inferred from the dashed curves on the right part of Fig.5.1, where instead on $T$ a dependence on the averaging interval is shown. The critical value of $\gamma_{\text{av}}$ is seen to lie between $1-2\ \hbar\omega_0$. Recalling that the typical nucleonic excitation occurs at $2\hbar\omega_0$, it is not astonishing that the *fluid of intrinsic particles* shows hydrodynamic behavior, but only for $\gamma_{\text{av}} \geq 2\hbar\omega_0$. Let us translate these values again into temperature by associating the $\gamma_{\text{av}}$ to the $\Gamma$ of (2.16) with $\hbar\omega = \mu$, or to the corresponding relaxation time from (3.5). Then for the critical temperature one gets values of $T \sim \sqrt{3\gamma_{\text{av}}}$, telling that for the nuclear fluid the transition to hydrodynamics occurs at the large temperatures of $T > \sqrt{3\hbar\omega_0}$. Actually this estimate is in accord with results of [18] obtained when describing the fluid of particles by a Landau equation with a collision term in relaxation time approximation. As can be seen from Fig.2 of this reference, the condition for collisional dominance in true sense, namely $\omega\tau \ll 1$, is fulfilled only at $T > 4$ MeV. For smaller temperatures nuclear modes do not behave like those of genuine hydrodynamics.

In *subsect.*{2.6.2} we learned that a behavior like $\propto T^{-2}$ may perhaps be seen at smaller temperatures as well, provided the contribution from the heat pole is to be taken seriously. Looking at (2.41) or (2.42) and recalling from Fig.2.4 that with increasing $T$ the factor $\chi^{\mathrm{T}} - \chi(0)$ quickly reaches a plateau, the decrease of $\gamma$ with $T$ above $T \approx 1.5$ MeV is to be attributed totally to the first factor being the nucleonic relaxation time. It is *only in this sense* that this behavior bears analogy with two-body viscosity. Otherwise it has not much do with it, as the factor $\chi^{\mathrm{T}} - \chi(0)$ essentially only reflects a dependency as given in the *independent particle model*. Indeed, in *subsect.*{2.6.1} we argued that for a situation as given by the nuclear compound model this contribution to friction from the heat pole should be expected to be much smaller. We claim the compound nucleus to show ergodic features which should ensure that the difference between adiabatic and isolated susceptibility must be small, if not identical to zero as in the ideal case supposed be (2.7). Then according to (2.38) the factor $\chi^{\mathrm{T}} - \chi(0)$ would reduce to $\chi^{\mathrm{T}} - \chi^{\mathrm{ad}}$, which for the present model would even vanish identically.

In *chap.*{3} we have seen that for $\omega\tau \ll 1$ a friction force of the type just mentioned also shows up in the model DDD. Indeed, we have seen (3.7) to be identical to the heat pole component (2.41) or (2.42), as it comes out within the independent particle model. Whether such configurations are important in actual situations, or whether one rather needs to refer more to the ones of the compound nucleus will largely be a question of time scales. Perhaps it is only for fission processes across a sufficiently high barrier that the system moves slowly enough such that typical compound configurations are reached. The model of DDD we described in *sect.*{3} was originally coined for the entrance phase of heavy ion collisions [15]. In such a situation the system may prefer the "diabatic"-like configurations such that the induced friction force may resemble more the one given by heat pole contribution.

In this context we want to mention once more the relaxation time approach to the time dependent shell model proposed in [17]. Also there two components of the friction force have been identified in eq.(29) as coming from contributions of diagonal or non-diagonal matrix elements. Because of the special property (2.35) given for the one body operator $\hat{F}$ whose susceptibilities we need to calculate, we realize that for the shell model such a separation is identical to the one into a heat pole plus the rest. In a forthcoming paper [45] numerical comparisons will be published. Qualitatively, the numerical results for friction presented there agree with those shown in Fig.5.4.

Finally, we like to come back to the paper [19] mentioned in the introduction. The friction force deduced there also shows the $T^{-2}$ behavior. Since the methods used are quite different from the ones we have developed, we have not been able to establish any closer formal contact. Nevertheless, we believe that the essential features of this work resemble more the ones of the compound nucleus rather than the ones of single particle picture, in particular not the ones of diabatic motion. We have essentially two reasons for this believe: i) as mentioned already in the introduction, in this model collisions play an essential role, ii) no diagonal matrix elements of the relevant one body operator contribute *.

## 6 Summary, conclusions and outlook

In this paper we have often come across differences between "diabatic" and "adiabatic" motion. The first case studied in *chap.*{2} was the effective coupling constant $k$ which appears in the strength distribution, or more generally in the collective response function (2.8). As seen from (2.18) and (2.19), an important contribution to $k$ is the stiffness of the static energy. In the "adiabatic" picture the latter is to be identified as the energy of the (quasi-static) equilibrium distribution, where in the ideal case assumed in (2.18) one ought to choose the internal energy $E(Q, S_0)$ at given entropy. In the pure "diabatic" picture the corresponding energy is the $E_{\text{di}}(Q)$ which is to be calculated for frozen occupation numbers. For details we may refer to [20]. There the temperature dependence of the two coupling constants have been compared with one another. Whereas $k_{\text{di}}^{-1}$ practically does not change with $T$ the "adiabatic" one strongly decreases with $T$. This is the main reason why the strength distribution of the "adiabatic" case changes so dramatically with $T$ [14]. We have been able to demonstrate this fact by showing the strength distribution for the pure "diabatic" case, for which such a shift in strength is absent and whose behavior is seen to be similar to the one generally expected for RPA.

In *chap.*{3} we have discussed a linearized version of the model of DDD [16], which allowed us to study a strength distribution for this theory as well. As the latter allows a transition or relaxation to the "adiabatic" surface, it was not astonishing to see this strength function having a similar feature to the one associated before to the "adiabatic" case. As mentioned in this context, such a behavior has also been seen in [18] within the framework of a special Landau-Vlasov approach to nuclear surface modes. In both cases the transition temperature was seen to be higher than for the one of the "adiabatic" case of [14] and [20]. More and even larger differences are found in transport coefficients, especially that of friction. We argued that the latter is suited best to distinguish between the various possible alternatives of describing nucleonic dynamics. We consider it one of the benefits of the present work that it allows one to study within one and the same framework such divers pictures as independent particle motion or the one of collisional dominance.

Before we continue to elaborate on this issue it may be worth while to briefly touch upon the question of why, for nuclear collective motion, there is dissipation at all, even in the form of a linear friction force. After all, the nucleus is a quantal micro-system. In the terminology of [46] the picture underlying our treatment is the one of "half-classical mechanics", as the nucleons are treated fully quantal and the collective variable appears as a classical c-number parameter. The reason why we find "quantal dissipation", nevertheless, is found in the experimental situation we want to describe. There is so much excitation energy in the system that the spectrum must be considered dense by all practical means. Using this argument one does not even need to invoke the fact that in true sense the nucleus is an open system having a finite decay width $\Gamma^\uparrow$. The experimental uncertainty in energy already permits one to apply smoothing procedures to a perhaps discrete spectrum, as given within a certain model. Indeed, such smoothing procedures were seen to be necessary in $chap.\{4\}$ to get a finite friction force within the pure independent particle model, which under certain circumstances may then be represented by the wall formula as the appropriate macroscopic (thermodynamic) limit. However, we have given arguments that, for a realistic nuclear situation, one needs to include configurations beyond the independent particle model. Here collisions are assumed to simulate such effects. When estimated through the imaginary part of the self-energy or the optical potential, the mean free path is seen to become smaller than the nuclear diameter. This is just another phrase of saying it is the more complex configurations of the compound nucleus, rather than the simpler ones of the pure shell model which matter. Back to energy averages, for the densely lying compound states the spacing is so small that such averages need not even be considered explicitly. They are implicit whenever one chooses to use smooth imaginary parts of the self-energies, as done in our case by way of the form (2.16).

---

* private communication by G.Bertsch

After these remarks of a more principle nature let us continue to compare the results for the friction coefficient which the various pictures deliver. Let us begin looking at the friction force (3.7) that the model of DDD predicts for slow motion, the only case we address to in the present work. It shows "hydrodynamical" behavior in the sense of being proportional to the relaxation time for single particle motion, which in turn is given by the inverse of the single particle width, typical for motion around the Fermi energy. For not too small temperatures this friction force behaves like $T^{-2}$. There are indications [2] that for the intermediate temperatures encountered in experiments, the absolute value of such a friction force is too large. This can be inferred from Fig.12 and 13 of [2] where microscopic studies are confronted with solutions of macroscopic equations and their comparisons with results of fission experiments (with references given therein). We have perhaps been able to find possible reasons for this discrepancy. This came up by interpreting the second factor in this friction coefficient in terms of thermodynamic concepts. This factor was seen to be proportional to the difference of the isothermal susceptibility and the static response, $\chi^T - \chi(0)$. We identified this form to represent that component of friction coming from the heat pole, as seen within linear response theory. If evaluated within an independent particle model, this difference $\chi^T - \chi(0)$ turns out large. The ultimate reason for this feature was seen to reflect violation of ergodicity, in the sense of having the adiabatic susceptibility be identical to the static response: $\chi^{ad} = \chi(0)$. Since for the nuclear case the difference between the isothermal and adiabatic susceptibility $\chi^T - \chi^{ad}$ is known to be small, ergodicity reduces this heat pole friction considerably. As a matter of fact, it would even decrease with increasing $T$, simply because $\chi^T - \chi^{ad}$ is known to decrease with $T$ [20].

The heat pole manifests itself in the correlation function as a peak at $\omega = 0$, which is nicely seen in Fig.2.2. Within our model, its width is approximately given by twice the single particle width calculated at the Fermi energy $\Gamma_T \approx 2\Gamma(\mu, T)$. It is worth recalling under what conditions ergodicity would be given, which is to say under which conditions the strength of the heat pole can be expected to reduce from $T(\chi^T - \chi(0))$ to $T(\chi^T - \chi^{ad})$. Following [22] it suffices to have a non-degenerate spectrum, if only the distribution in energy is sufficiently narrow. One is inclined to anticipate such a situation to be given for the compound nucleus, with a correct treatment of the equilibrium density being based on the microcanonical ensemble. Its level spacing is known to follow a Wigner distribution rather than the one of Poisson (for a discussion of the complication due to a few conserved quantities see *subsect.*{2.6.1}). Indeed, model studies based on an evaluation of response and correlation functions within a Random Matrix Model show no traces of a heat pole. The correlation function presented in the appendix of [47] has no peak at $\omega = 0$ (see also forthcoming publications).

Apparently, and perhaps unfortunately, collisional damping as used in the present paper does not lead to ergodic behavior. As compared to the pure shell model, the entire strength in the heat pole does not change when collisions are turned on. There could be several reasons for this fact. First of all, as mentioned in the text, the renormalized single particle energies show the same degeneracies as the ones of the original shell model. This deficiency could be cured by using state dependent self-energies. There is a second point which in some sense is related to the first. The deformed shell model delivers "diabatic" states, but ergodicity is favored by "adiabatic" ones (see the discussion in part a) of *subsect.*{2.6.1}). Such a transition from a "diabatic" to an "adiabatic" spectrum is not included in the present treatment. It would require to directly couple two states when they come close. This effect would certainly influence the incoherent excitations at very small frequencies. Another deficiency may be found in the fact that, for pragmatic reasons, we are forced to use the canonical distribution when computing the response or correlation function. Indeed, as mentioned in *subsect.*{2.6.1}, here one is confronted with a situation in which it is the fluctuation of the energy distribution itself, which matters. Finally, one should not forget that in the present work a case of particularly high symmetry has been taken. It may just be that the strength of the heat pole becomes abnormally large for the spherical shape. Indeed, for fission such a configuration is a very singular one. The shapes the system passes through on its way to scission are of much higher complexity. The importance of this feature in relation to the degree of "chaoticity" of nuclear dynamics has been pointed out by W.J. Swiatecki [48]. There can perhaps be little doubt that for the present problem chaos and ergodicity are related to each other in one way or another. Curiously enough, as seen before, ergodicity *reduces* the friction coming from the heat pole rather than increasing it it.

To ensure ergodicity we suggested to just replace in the Lorentzian corresponding to the heat pole the $\chi^{\rm T} - \chi(0)$ by $\chi^{\rm T} - \chi^{\rm ad}$. As for the present model the latter difference is identically zero, we argued to simulate ergodic behavior by leaving out the contribution coming from diagonal matrix elements. Results of such computations were shown in *sect.{5.2}*. As discussed there and shown in Fig.5.2 the temperature dependence of this component depends on the detailed form of the self-energy as a function of frequency and temperature. For a finite cut-off parameter $c$ friction has the tendency to first increase with $T$ and then to level off around $T \approx 4$ MeV. The value it reaches is somewhat smaller than the one of the wall formula for an infinite system. Conversely, if the cut-off is discarded friction decreases again at large temperatures, showing in this sense signs of collisional damping; for details please see the text.

The contribution of the heat pole has been left out also in the previous computations [11], [12], [14] in which the transport coefficients were deduced in a self-consistent fashion, where one is not forced to rely on the validity of the zero-frequency limit. In this method one adjusts a peak in the collective strength function, the dissipative part of $\chi_{\rm coll}(\omega)$ as given in (2.8), to the one of the oscillator response shown in (2.20). From the structure of $\chi_{\rm coll}(\omega)$ it is easy to understand that such a fit could as well be done for the nucleonic response function $\chi(\omega)$ (cf.[32]). In case the heat pole is important one would have to generalize to a fit with two Lorentzians, which in such a context has as yet not been undertaken.

Let us finally address once more the question of possible experimental verifications. We mentioned indications that the friction coming from the heat pole may lead to too large values. On the other hand, the authors of [3] and [4] have opted for a friction coefficient which, as function of the excitation energy, sets in quite sharply at small temperatures. Could this hint at the importance of the heat pole friction? We are certainly not able to settle this question at the moment. This requires more careful studies, both of microscopic nature as well as of those involving solutions of macroscopic equations and their comparisons with experimental results. But we like to finish by adding the following warning remarks. First of all, in the regime of small temperatures pairing correlations cannot be neglected. Their influence on the friction coefficient is expected to result in a similar behavior: Large pairing will reduce the imaginary part of the self-energies [49] and hence collective dissipation at small $T$. Secondly, from the arguments given above about the independent particle model overestimating the contribution from the heat pole, one is inclined to argue that such a sharp rise, as is seen for such a model, could only be realized for fission processes which occur on some intermediate time-scale. By this we want to stress that for a situation in which the system sees a large fission barrier, it will have enough time to explore all the many compound configurations for which one may safely assume ergodicity to be given.

### Acknowledgements

The authors like to express their gratitude for useful suggestions obtained in discussions with G. Bertsch, W. Brenig, A.S. Jensen, R. Lemmer and W. Nörenberg. Especially, they want to thank D. Kiderlen whose critical comments on preliminary versions of this paper have helped considerably to straighten the lines of arguments about the thermodynamic problems. Furthermore, we like to acknowledge financial support by the DFG and two of us (F.A.I. and S.Y.) want to thank the Physik Department of the TUM for the hospitality extended to them during their stay.

## A  Appendices

### A.1  Single particle models

For spheroidal isoscalar quadrupole vibrations around a spherical equilibrium the operator $\hat{F}$ is given by

$$\hat{F}(\hat{x}_i, \hat{p}_i) = \frac{\partial \hat{H}(\hat{x}_i, \hat{p}_i, Q)}{\partial Q}\bigg|_{Q=0} = \frac{\partial V(\vec{r}, Q)}{\partial Q}\bigg|_{Q=0} = \frac{\partial V(r, Q)}{\partial Q}\bigg|_{Q=0} Y_{20}(\cos\theta) \quad (A.1)$$

provided for a spherical nucleus the state $|k\rangle$ is specified by the main quantum number $n$, orbital momentum $l$ and its projection $m$ and the matrix elements of $\hat{F}$ is the product radial and angular integrals, $\left|F_{kj}\right|^2 = F^2_{nl,n'l'} h_{ll'}$ where

$$F^2_{nl,n'l'} = \left|\langle nl|\frac{\partial V(r,Q)}{\partial Q}|n'l'\rangle\right|^2 \qquad \text{and} \qquad h_{ll'} = \sum_{mm'} |\langle Y_{lm}|Y_{20}|Y_{l'm'}\rangle|^2 \quad (A.2)$$

The angular matrix elements (2) obey obvious selection rules, namely $h_{ll'} \neq 0$ if $l' = l, l \pm 2$. Due to these rules the double sum in (4.17) gets reduced to just one sum over $l$. Further derivations depend on the specific properties of the radial matrix elements $F^2_{nl,n'l'}$. Therefore, it is more convenient to proceed separately for the infinitely deep square well with its sharp edge and for the oscillator potential.

### A.1.1 The infinitely deep square well

To some extent an infinitely deep potential may simulate properties of a finite nucleus. Neglecting spin orbit forces calculations can largely be done analytically. The drawback is that the (nucleonic) response function we like to have in our formulation shows some peculiar behavior. This is due to the fact that for such a potential the field $F$ is "surface peaked" to the extreme of a $\delta$ function. This implies that for the matrix elements there is no natural cut off. The cut off parameter $E_{\text{cut}}$ was introduced here in order to make the response functions decrease with increasing $\omega$. But doing this, interesting features can be deduced.

The potential is defined as

$$V(\mathbf{r}, Q) = -V_0 \Theta(r - R(\theta, Q)), \quad \text{with} \quad V_0 \to \infty \quad (A.3)$$

and

$$R(\theta, Q) = R_0 \left(1 + QY_{20}(\cos \theta)\right) \quad (A.4)$$

Such a potential requires special boundary conditions. We use the same prescription as [43], namely

$$\lim_{V_0 \to \infty} \frac{\sqrt{2mV_0}}{\hbar} \mathcal{R}_{nl}(R_0) = \left. \frac{\partial \mathcal{R}_{nl}(r)}{\partial r} \right|_{r=R_0} \quad (A.5)$$

with $\mathcal{R}_{nl}(r)$ being the radial wave functions of the square well potential. The matrix elements $F^2_{nl,n'l'}$ of the radial part of $\hat{F}$

$$\left. \frac{\partial V(r, Q)}{\partial Q} \right|_{Q=0} = V_0 R_0 \delta(r - R_0) \quad (A.6)$$

turn out separable in $nl$ and $n'l'$ and can be written as

$$F^2_{nl,n'l'} = 4\epsilon_{nl}\epsilon_{n'l'} \quad (A.7)$$

where the energies $\epsilon_{nl}$ are the eigenvalues of the Schrödinger equation for the potential (3).

### A.1.2 The oscillator potential

For spheroidal deformations of a harmonic oscillator without spin orbit couplings

$$V_{osc}(\epsilon, \vec{r}) = \frac{m}{2}(\omega_\perp^2(x^2 + y^2) + \omega_z^2 z^2) \quad (A.8)$$

with deformation parameter $\epsilon$ defined in the same way as in Nilsson model [50]

$$\frac{\omega_\perp}{\omega_z} = \frac{1 + \epsilon/3}{1 - 2\epsilon/3} \quad (A.9)$$

friction has been evaluated similar to the case of square well potential.

Differentiating (8) with respect to $\epsilon$ one gets the $\hat{F}$ operator

$$\hat{F}(\epsilon, \vec{r}) = \frac{\partial V_{osc}}{\partial \epsilon}\bigg|_{\epsilon=0} = -\frac{2}{3}\sqrt{\frac{4\pi}{5}} m\omega_0^2 r^2 Y_{20}(\cos\theta) \qquad (A.10)$$

For a spherical shape the matrix elements of $\hat{F}$ are separated into the angular and radial parts. The angular part is just the same as for the square well potential and radial part is

$$F_{nl,n'l'}^2 = \frac{16\pi}{45} m^2 \omega_0^4 |\langle nl|r^2|n'l'\rangle|^2 \qquad (A.11)$$

Contrary to the case of square well potential where matrix elements $F_{nl,n'l'}^2$ were different from zero for any $n$ and $n'$ for oscillator potential $F_{nl,n'l'}^2$ are not zero only if $n' = n \pm 1$, namely [50]

$$\begin{aligned}\langle N-2, l|\xi^2|N, l\rangle &= \sqrt{n(n+l+1/2)} \\ \langle N-2, l-2|\xi^2|N, l\rangle &= \sqrt{(n+l+1/2)(n+l-1/2)} \\ \langle N-2, l+2|\xi^2|N, l\rangle &= \sqrt{n(n-1)}\end{aligned} \qquad (A.12)$$

In (12) $N$ is the mean quantum number, $N = 2n + l$, defining the single-particle energies

$$\epsilon_{nl} = (2n + l + 3/2)\hbar\omega_0 \qquad (A.13)$$

and

$$\xi^2 = m\omega_0 r^2/\hbar \qquad (A.14)$$

Besides the matrix elements (12) the diagonal in $N$ matrix elements

$$\begin{aligned}\langle N, l|\xi^2|N, l\rangle &= N + 3/2 \\ \langle N, l-2|\xi^2|N, l\rangle &= 2\sqrt{(n+1)(n+l+1/2)}\end{aligned} \qquad (A.15)$$

also different from zero. These matrix elements do not contribute to the frequency or density smoothed response function. But they do contribute to the spectral smoothed response function and this is the main contribution to the friction coefficient at large averaging parameter $\gamma_{av} \geq 2\hbar\omega_0$.

## A.2 The wall formula

In the classic paper [5] the formula for energy dissipation is given as

$$\dot{E} = \frac{3}{4}\rho v_F \oint u_n^2(s) ds \qquad (A.16)$$

Here $\dot{E}$ is the energy transferred to the gas by moving surface and $u_n$ is the normal velocity of the surface. For the small deviations from the spherical shape (A.4) considered here the normal velocity is

$$u_n = \frac{\partial R(\theta, t)}{\partial t} = R_0 Y_{20}(\cos\theta)\dot{Q}(t) \qquad (A.17)$$

and the surface integral in (16) is equal to $R_0^4$. With obvious notations

$$\rho = mA / \frac{4}{3}\pi R_0^3, \quad v_F = \frac{\hbar k_F}{m} = \frac{\hbar x_F}{m R_0} \qquad (A.18)$$

the formula (16) becomes

$$\dot{E} = \gamma_w \dot{Q}^2, \qquad \gamma_w = \frac{9A\hbar}{16\pi} x_F \qquad (A.19)$$

The numerical value of (4.24) and (19) depend essentially on the Fermi momentum $k_F = x_F/R_0$. If one relates Fermi momentum to the number of particles by Thomas-Fermi approximation

$$A(\epsilon_F) = \frac{2}{9\pi} x_F^3 \qquad (A.20)$$

then the friction coefficient given in (19) and (4.24) coincide with each other. However in the calculations with finite nuclei one usually relates Fermi momentum (or Fermi energy) to the number of particles by integration the (smoothed) density of single-particle states,

$$A = \int^{\epsilon_F} \bar{g}(e) de, \quad \epsilon_F = \frac{\hbar^2 x_F^2}{2m R_0^2} \qquad (A.21)$$

with the smoothed density defined by (4.10). For the comparison of friction coefficients (19) and (4.24) with $\bar{\gamma}$ defined by Strutinsky smoothing it seems more meaningful to use the value $x_F$ (21) since it appears in the calculations of averaged quantum quantities like averaged response and $\bar{\gamma}$. However, substitution of this value of Fermi momentum into the formulae (4.24) or (19) gives different values for friction. These two different wall formula values are shown by the lines with stars in Fig.5.1. The upper one is defined by equations (4.24), (21) and lower one corresponds to (19),(21). In the preliminary version of this paper [51] we have shown only upper one as the wall formula value.

**Figure captions**

Fig.2.1 Strength distribution for diabatic quadrupole vibrations

Fig.2.2 Imaginary parts of correlation ($\psi''$; upper part) and response functions ($\chi''$; lower part), for two temperatures. For $\psi''$ the dashed curve shows the "heat pole", for $\chi''$ it shows the result when the contribution from the latter is removed.

Fig.2.3 The width $\Gamma_T$ of the heat pole as function of temperature (see text).

Fig.2.4 $\psi^0/T = \chi^T - \chi(0)$ as function of temperature; solid curve for collisional damping, the dashed one for independent particle motion.

Fig.2.5 Contribution of the "heat pole" to friction, for the "non-ergodic" system: for the fully drawn curve the $\Gamma(\mu,T)$ is evaluated for $c = 20$ MeV, and for dashed curve for $1/c = 0$. As reference values we indicate the result of the wall formula (line with stars) and show the contribution from the non-diagonal matrix elements (line with squares).

Fig.3.1 The imaginary part of $\chi_{coll}(\omega)$ scaled with the inertia for irrotational flow, $B$, for several temperatures $T$.

Fig.4.1 The smoothed density of single-particle states with fixed orbital momentum $l$ (indicated by numbers) divided by the corresponding density in Thomas-Fermi approximation as function of the averaging parameter $\Delta$, see text. The solid curve with stars shows the full density of single-particle states divided by the same quantity in Thomas-Fermi approximation. The computations were carried out for the single-particle spectrum of infinitely deep square well potential. The Fermi momentum $k_F$ corresponds to a system with $A = 138$ particles.

Fig.5.1 Left hand side: The averaged friction coefficient (4.18) as function of the averaging parameter $\Delta$. The different solid curves refer to different degrees of Strutinsky's smoothing function (4.3). The lines with stars correspond to the two versions of wall formula explained in the Appendix A.2. The single-particle spectrum of square well potential was cut at $E_{\rm cut} = 30 MeV$ (top) and at $E_{\rm cut} = 100 MeV$ (bottom) above Fermi energy. The number of particles is $A = 138$.

Right hand side: The friction coefficient (4.17) computed for $A = 112$ particles in a harmonic oscillator potential. The averaging interval parameter $\Gamma/2$ is measured in units of the inter-shell distance $\hbar\omega_0$. The lines with stars correspond to the value of the wall formula (A.19),(5.1). The dashed curves are obtained by smoothing with a Lorentzian.

Fig.5.2 The temperature dependence of the friction coefficient (2.24) in zero frequency limit computed with the square well (left, number of particles $A = 138$) and oscillator (right, $A = 112$) potentials and assuming ergodicity in the sense of (2.7). The dashed curves are obtained within the "on shell" approximation.

Fig.5.3 The friction coefficient calculated from the response function including the contributions from the "heat pole", with the collisional widths calculated in the following way: Fully drawn line: $\Gamma(\hbar\omega, T)$ as given by (2.16); short-dashed curve: $\Gamma(\hbar\omega = \mu, T)$, long-dashed curve: on-shell approximation, i.e. $\Gamma(\hbar\omega = e_k, T)$. The contribution from the non-diagonal matrix elements and the value of the wall formula are shown in addition.